\documentclass[twocolumn,trackchanges]{aastex6}

\begin{document}

\title{Deep Subaru Hyper Suprime-Cam observations of Milky Way satellites Columba I and Triangulum II \footnotemark[1]} \footnotetext[1]{Based in part on data collected at Subaru Telescope, which is operated by the National Astronomical Observatory of Japan.}

\author{Jeffrey L. Carlin\altaffilmark{2}}
\altaffiltext{2}{LSST, 950 North Cherry Avenue, Tucson, AZ 85719, USA; jcarlin@lsst.org} 

\author{David J. Sand\altaffilmark{3,4}}
\altaffiltext{3}{Texas Tech University, Physics \& Astronomy Department, Box 41051, Lubbock, TX 79409-1051, USA}
\altaffiltext{4}{Department of Astronomy/Steward Observatory, 933 North Cherry Avenue, Rm. N204, Tucson, AZ 85721-0065, USA}

\author{Ricardo R. Mu\~noz\altaffilmark{5}}
\altaffiltext{5}{Departamento de Astronom\'ia, Universidad de Chile, Casilla 36-D, Santiago, Chile}

\author{Kristine Spekkens\altaffilmark{6,7}}
\altaffiltext{6}{Department of Physics, Engineering Physics and Astronomy, Queen's University, Kingston, Ontario, Canada, K7L 3N6}
\altaffiltext{7}{Department of Physics, Royal Military College of Canada, P.O. Box 17000, Station Forces, Kingston, ON K7L 7B4, Canada}

\author{Beth Willman\altaffilmark{8}}
\altaffiltext{8}{LSST and Steward Observatory, 933 North Cherry Avenue, Tucson, AZ 85721, USA} 

\author{Denija Crnojevi{\'c}\altaffilmark{3}}

\author{Duncan A. Forbes\altaffilmark{9}}
\altaffiltext{9}{Centre for Astrophysics and Supercomputing, Swinburne University, Hawthorn VIC 3122, Australia}

\author{Jonathan Hargis\altaffilmark{10}}
\altaffiltext{10}{Space Telescope Science Institute, 3700 San Martin Drive, Baltimore, MD 21218, USA}

\author{Evan Kirby\altaffilmark{11}}
\altaffiltext{11}{California Institute of Technology, 1200 E. California Boulevard, MC 249-17, Pasadena, CA 91125, USA}

\author{Annika H. G. Peter\altaffilmark{12}}
\altaffiltext{12}{CCAPP, Department of Physics, and Department of Astronomy, The Ohio State University, Columbus, OH 43210, USA}

\author{Aaron J. Romanowsky\altaffilmark{13,14}}
\altaffiltext{13}{Department of Physics and Astronomy, San Jos\'e State University, One Washington Square, San Jos\'e, CA 95192, USA}
\altaffiltext{14}{University of California Observatories, 1156 High Street, Santa Cruz, CA 95064, USA}

\author{Jay Strader\altaffilmark{15}}
\altaffiltext{15}{Department of Physics and Astronomy, Michigan State University, East Lansing, MI 48824, USA}

\begin{abstract}

We present deep, wide-field Subaru Hyper Suprime-Cam photometry of two recently discovered satellites of the Milky Way (MW): Columba~I and Triangulum~II.  The color magnitude diagrams of both objects point to exclusively old and metal-poor stellar populations. We re-derive structural parameters and luminosities of these satellites, and find $M_{\rm V, Col~I} = -4.2\pm0.2$ for Col~I and $M_{\rm V, Tri~II} = -1.2\pm0.4$ for Tri~II, with corresponding half-light radii of $r_{\rm h, Col~I} = 117\pm17$~pc and $r_{\rm h, Tri~II} = 21\pm4$~pc. The properties of both systems are consistent with observed scaling relations for MW dwarf galaxies. Based on archival data, we derive upper limits on the neutral gas content of these dwarfs, and find that they lack H\textsc{I}, as do the majority of observed satellites within the MW virial radius. Neither satellite shows evidence of tidal stripping in the form of extensions or distortions in matched-filter stellar density maps or surface density profiles. However, the smaller Tri~II system is relatively metal-rich for its luminosity (compared to other MW satellites), possibly because it has been tidally stripped. Through a suite of orbit simulations, we show that Tri~II is approaching pericenter of its eccentric orbit, a stage at which tidal debris is unlikely to be seen. In addition, we find that Tri~II may be on its first infall into the MW, which helps explain its unique properties among MW dwarfs. Further evidence that Tri~II is likely an ultra-faint dwarf comes from its stellar mass function, which is similar to those of other MW dwarfs.

\end{abstract}


\section{Introduction} \label{sec:intro}

The number of known Milky Way satellites has been increasing rapidly over the past several years due to the availability of large-area, deep, high-precision photometric catalogs from imaging surveys such as the Sloan Digital Sky Surveys \citep{York2000_SDSS,SDSSII,SDSSIII}, ATLAS \citep[][]{Shanks15_ATLAS}, the Panoramic Survey Telescope \& Rapid Response System \citep[Pan-STARRS;][]{ChambersPS1}, the Dark Energy Survey \citep[DES;][]{DES05,DES16} and other surveys employing the Dark Energy Camera (e.g., MagLiteS: \citealt{Drlica16,Drlica-Wagner17_MagLiteS}; SMASH: \citealt{Nidever17_SMASH}).
Many of these new discoveries fall into the ``ultra-faint dwarf (UFD)'' category \citep[e.g.,][]{Willman05_UMa,Belokurov06_Bootes,Zucker06_Uma,Zucker06_CVn,Belokurov07_5MWsats,Walsh07_BooII,Bechtol15,Kim15a,Kim15b,Koposov15,Laevens15,Martin15,Drlica16,Torrealba16}, with luminosities as low as a few hundred $L_\odot$.  The UFDs are apparently the most dark matter (DM) dominated systems known \citep[e.g.,][]{Simon07}, though the so-called ``ultra-diffuse galaxies'' may have extremely high DM fractions as well \citep[e.g.,][]{koda15,beasley16,vanDokkum16,zaritsky17}.
Because there are UFDs that are fainter than (and as small as) bright globular clusters (GCs), the line between star clusters and dwarf galaxies has become blurred. \citet{WillmanStrader12} argued that a ``galaxy'' should be defined as an object whose properties cannot be readily explained by a combination of Newtonian gravity and baryons, whereas \citet{forbes_kroupa2011} advocated a dynamical criterion and/or the presence of complex stellar populations to define a galaxy. Evidence that the UFDs differ from GCs in ways that satisfy these definitions of a galaxy is seen in the form of metallicity spreads among their stars (such that they must have had deep potential wells to retain their gas for extended periods) and large velocity dispersions (suggesting their kinematics are dominated by dark matter).

As the most dark matter-dominated, chemically pristine objects in the Universe \citep[e.g.,][]{Munoz06_Bootes,Brown12,Kirby13}, the UFDs are important laboratories in which to seek clues to the nature of dark matter (via, e.g., searching for gamma-ray signals due to DM particle annihilation; \citealt{Albert17}).
Given that tidal forces are likely to have shaped their properties \citep[e.g.,][]{penarrubia08}, it is also curious that Local Group dwarfs follow well-determined scaling relations \citep[for several examples, see][]{McConnachie12}. \citet{Munoz12} showed that the structural parameters (e.g., size, stellar density, luminosity) of UFDs derived using small numbers of stars are biased, so it is important that we measure their properties via deep imaging and precise photometry. With imaging over a sufficient area around a given dwarf, one can also search for signs of tidal disruption in the stellar density \citep[e.g.,][]{Sand09,Munoz10,Sand12}.  Deep follow-up photometry of the new MW satellites has already led to other surprises, including the smallest galaxy known to host its own star cluster in Eridanus II \citep{Crnojevic16}, which has yielded its own constraints on dark matter properties \citep[e.g.,][]{Brandt16,Amorisco17,Contenta17_EriII}.

In this contribution, we detail our deep Subaru+Hyper Suprime-Cam imaging around two recently discovered UFD candidates -- Columba~I (Col~I) and Triangulum~II (Tri~II). Col~I is a distant ($d\sim180$~kpc) UFD candidate that was discovered as an overdensity of red giant branch (RGB) stars in DES Year 2 data \citep{dbr+15}. Its properties as measured by \citet{dbr+15} are fairly typical of Local Group UFDs, with a half-light radius of $\sim$100~pc and luminosity $\sim5\times10^3~L_\odot$ ($M_V\approx-4.5$). This candidate dwarf shows a prominent blue horizontal branch (BHB) and a sparsely populated RGB. Our aim in the current work is to derive structural parameters by probing nearly to the main sequence turnoff (MSTO) of Col~I, which provides a much larger sample of stars with which to robustly determine the properties of this distant, faint dwarf candidate. We also search for tidally-induced distortions in its outer regions.

Triangulum~II (Tri~II) was discovered by \citet{lmi+15} as a stellar excess at a distance of $\sim30$~kpc in the PanSTARRS1 database, and confirmed with deeper Large Binocular Telescope (LBT) imaging. Laevens et al. found Tri~II to be extremely faint ($M_{\rm V} \sim -1.8$, or $L \sim 450~L_\odot$) and compact ($r_{\rm h} \sim 30$~pc), with very few RGB stars, but apparently metal-poor stellar populations. There have been multiple spectroscopic follow-up programs of Tri~II \citep{kcs+15,mic+16,kcs+17,vsm+17}.  While there is some debate as to its dynamical status and dark matter content \citep[e.g.,][]{mic+16,kcs+17}, the apparent presence of a metallicity spread is indicative of a dwarf galaxy origin \citep{kcs+17}. Our extremely deep imaging may shed light on the equilibrium status of this system.


\begin{figure*}[!t]
\begin{center}
\includegraphics[width=0.45\textwidth, trim=0.0cm {0.0cm} {0.0cm} {0.0cm}, clip]{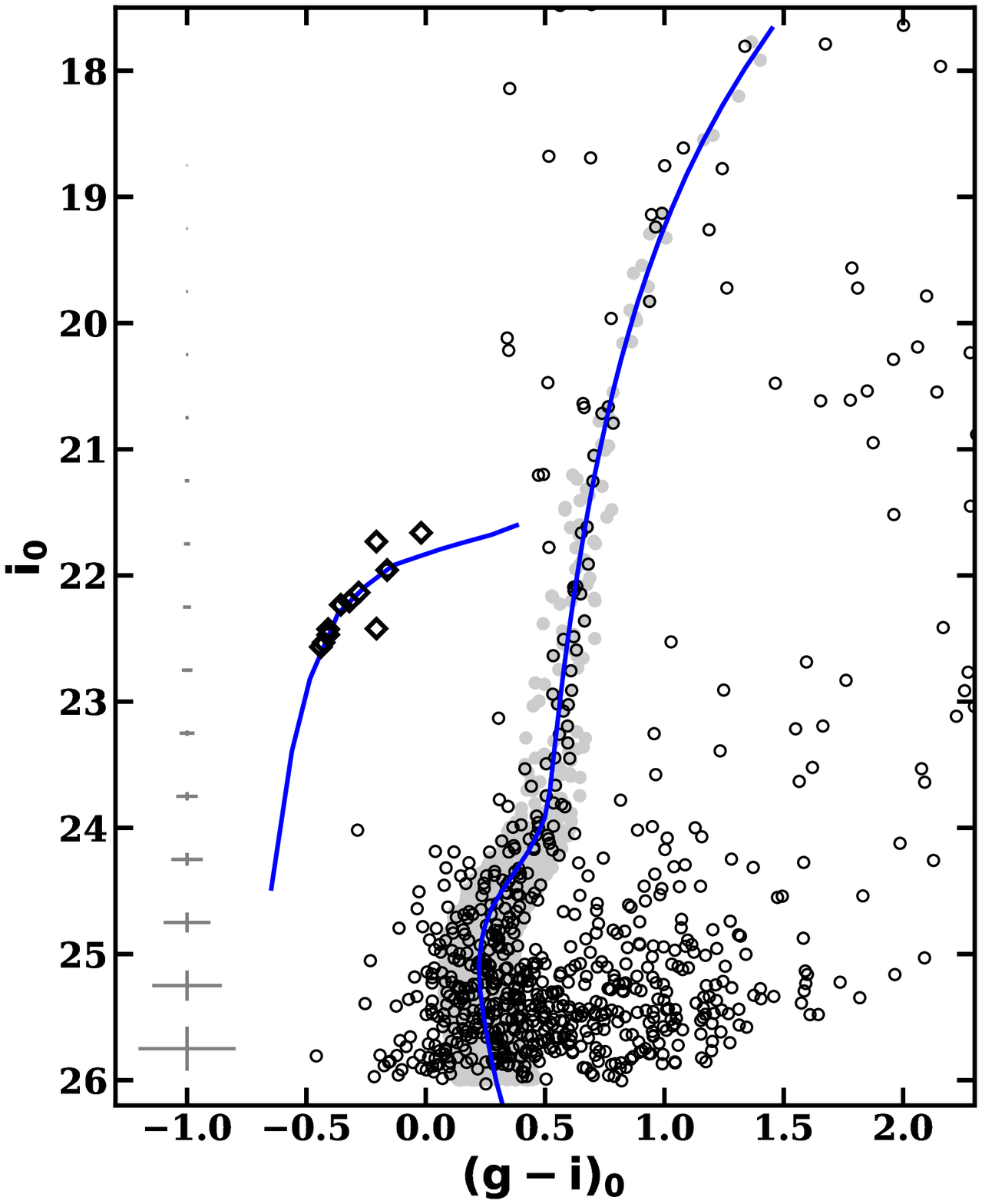}
\includegraphics[width=0.45\textwidth, trim=0.0cm {0.0cm} {0.0cm} {0.0cm}, clip]{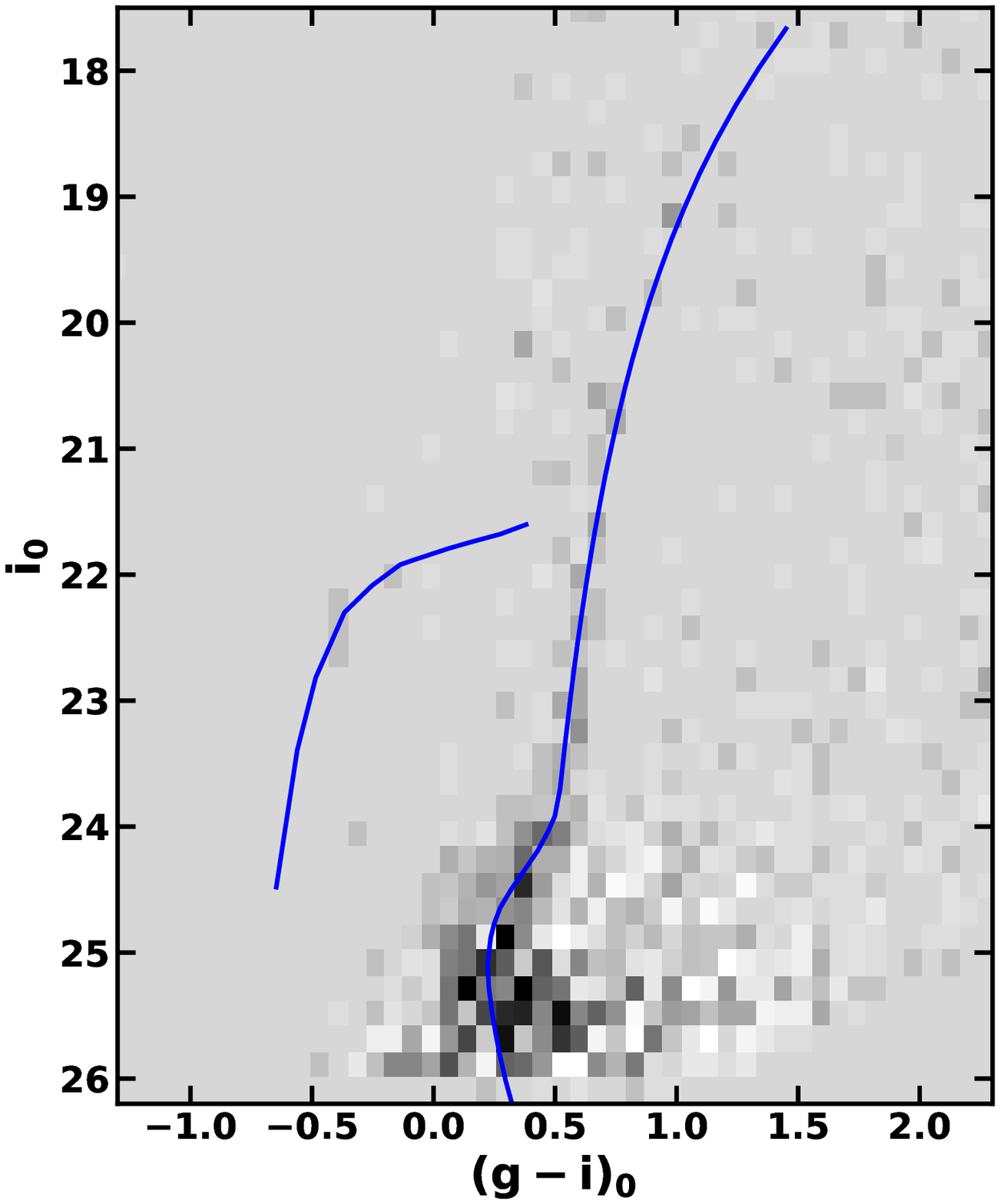}
\caption{{\it Left:} Color magnitude diagram showing (open circles) all well-measured (magnitude errors less than 0.2) point sources within $2.2'$ (our measured half-light radius) of the center of Col~I. Large open diamonds are BHB candidates, for which we include sources out to $5'$ from the center. Magnitudes have been calibrated to PanSTARRS public data \citep{ChambersPS1}. Large gray points are stars within $12'$ of the center of Col~I, selected by our isochrone filter, which constitute the sample used to derive structural parameters. Overplotted as a blue line is the PanSTARRS ridgeline for globular cluster NGC~7078 (M~15) from \citet{bfs+14}, shifted to a distance modulus of $m-M=21.31$ that we derived by fitting the M~15 BHB ridgeline to our Col~I data. 
Median photometric errors in 0.5-mag bins are shown along the left side of the figure. {\it Right:} CMD Hess diagram of the same field of view shown in the left panel, but with the average density of four equal-area background fields subtracted. The MSTO of Col~I stands out more clearly once the background is removed. The ridgeline of the old ($>10$~Gyr), metal-poor ([Fe/H]~$\sim$~-2.34) cluster NGC~7078 closely matches the stellar population of Col~I; hence, Col~I must also consist of a predominantly old, metal-poor population.
}
\label{fig:coli_CMD}
\end{center}
\end{figure*}

\begin{figure*}[!t]
\begin{center}
\includegraphics[width=0.45\textwidth, trim=0.0cm {0.0cm} {0.0cm} {0.0cm}, clip]{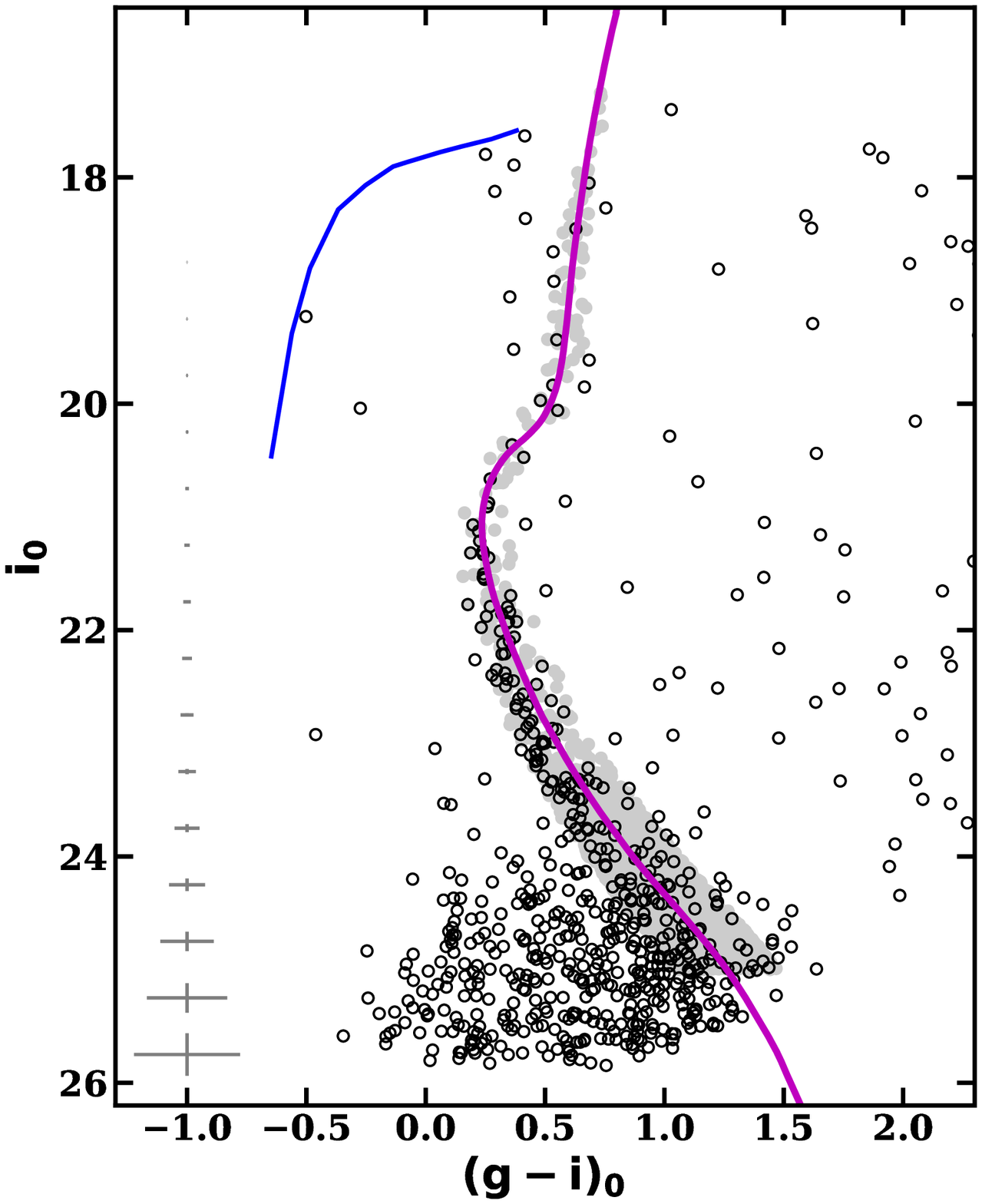}
\includegraphics[width=0.45\textwidth, trim=0.0cm {0.0cm} {0.0cm} {0.0cm}, clip]{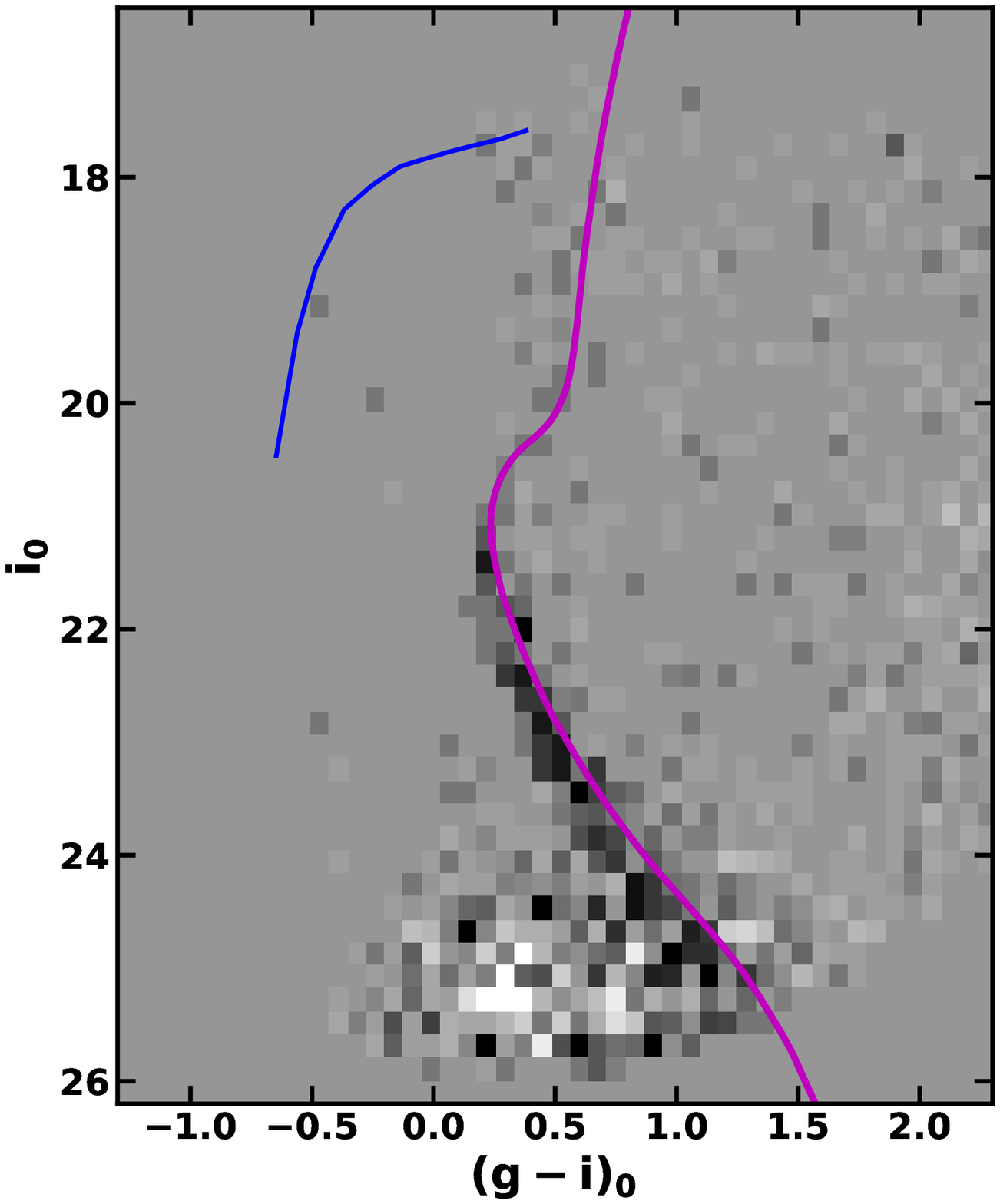}
\caption{{\it Left:} Color magnitude diagram of point sources within $2.5'$ (roughly our measured half-light radius) of the center of Tri~II as open circles. Magnitudes have been calibrated to PanSTARRS public data. As in Figure~\ref{fig:coli_CMD}, large gray points illustrate the sample selection for structural parameter derivation; for Tri~II this includes stars within $15'$ of the center. Overplotted as a blue line is the PanSTARRS BHB ridgeline for globular cluster NGC~7078 (M~15) from \citet{bfs+14}, shifted by a distance modulus of $m-M=17.27$ that we derived by fitting the M~15 main sequence ridgeline to our Tri~II data. Because the empirical M~15 ridgeline only traces the upper main sequence, we use a theoretical isochrone for Tri~II analysis. We have overlaid a Dartmouth isochrone \citep[magenta sequence; ][]{dotter08} with age 13 Gyr, [Fe/H] = -2.0, and $[\alpha$/Fe] = 0.4, shifted to the distance of Tri~II. This is the isochrone that we used to filter stars for deriving the structural parameters. Median photometric errors in 0.5-mag bins are shown along the left side of the figure. {\it Right:} CMD Hess diagram of the same field of view shown in the left panel, but with the average density of four equal-area background fields subtracted. The main sequence of Tri~II is well defined to $i_0 \sim 25$ once the background is removed. 
}
\label{fig:triii_CMD}
\end{center}
\end{figure*}

\section{Observations and Data Reduction} \label{sec:data}

We observed Col~I on 2016 Feb 10 and Tri~II on 2016 Feb 09 with Hyper Suprime-Cam \citep[HSC;][]{HSC} on the Subaru 8.2m telescope during scheduled time that was unusable for our Magellanic Analogs Dwarf Companions And Stellar Halos (MADCASH) Local Volume galaxies program \citep[see][]{Carlin16}. HSC has a 1.5$^{\circ}$ diameter field of view that easily encompasses the new MW satellites, which are typically a few arcminutes in size, and allows for a search for extended low surface brightness features that may have been missed in the discovery data. Seeing during the Tri~II observations was $\sim0.8-1.0''$, under photometric conditions. For the Col~I observations we had $\sim0.7-1.0''$ seeing, under clear, photometric skies. Exposure times were $12\times150$ sec in $i$ and $8\times300$ sec in $g$ for both Tri~II and Col~I, reaching $5\sigma$ limiting depths of $g \sim 26.9$ and $i \sim 25.8$ for Tri~II, and $g \sim 27.1$, $i \sim 25.9$ for Col~I in the reduced co-added frames.

The data were reduced using $hscPipe$ \citep{Bosch17_hscPipe}, a modified version of the LSST software stack \citep{IvezicLSST08,JuricLSSTDM15}, including standard processing steps, co-adding of the individual (dithered) frames, and PSF photometry.
Our final stellar catalog included all objects classified as ``point-like'' by the $hscPipe$ star/galaxy classifier, which is based on the difference between PSF and \verb cmodel \footnote{http://www.sdss3.org/dr10/algorithms/magnitudes.php\#cmodel}~magnitudes for each object (similar to \verb classification_extendedness; ~e.g., \citealt{Aihara17,Huang17}).

\subsection{Catalogs and Color-Magnitude Diagrams}\label{sec:cmds}

We calibrate the photometry by matching to the PanSTARRS (PS1) survey \citep{ChambersPS1,Flewelling16,Magnier16}, and fitting transformations in $g$ and $i$ that include both a magnitude and color dependence (including only stars between $18 < g < 21$ and $17.5 < i < 20.5$ in the fits).  We apply extinction corrections based on the \citet{schlafly11} modifications of the \citet{SFD98} reddening maps. The mean reddening values along the lines of sight to Col~I and Tri~II are $E(B-V) = 0.03$ and $0.07$, respectively. All photometry used throughout this work has been calibrated onto the PS1 system and corrected for line-of-sight extinction.

We present the final color magnitude diagrams (CMDs) of Col~I in Figure~\ref{fig:coli_CMD} and Tri~II in Figure~\ref{fig:triii_CMD}. 
For Col~I's CMD, we include stars within $2.2'$ (our measured half-light radius; see Section~\ref{sec:struct_params}) of the dwarf's center. Candidate blue horizontal-branch stars (BHBs) within $5'$ of the Col~I center are displayed as large open diamonds, highlighting the prominent BHB of this galaxy. Col~I shows a narrow and well-defined red giant branch (RGB), and a clear main sequence turnoff (MSTO) that mingles with unresolved background galaxies at magnitudes $i_0 \gtrsim 25.5$. The right panel of Figure~\ref{fig:coli_CMD} encodes the number density of stars in the same region as the left panel, after subtracting the average density in each color/mag bin from four equal-area background fields well outside the body of Col~I. The MSTO is much clearer in this Hess diagram, which accounts for the average number of contaminating background galaxies. 

Figure~\ref{fig:triii_CMD} shows the CMD of stars within $2.5'$ of Tri~II, which reaches $\gtrsim4$~mags below the MSTO. The main sequence is narrow and well-defined, with a sparsely populated RGB and no evidence for a BHB population. (Note that the bright end of the RGB is cut off because our deep data saturate at $i_0 \sim 18$.) In the background-subtracted Hess diagram (right panel of Fig.~\ref{fig:triii_CMD}), the main sequence is clear down to the limiting magnitude of our data at $i_0 \gtrsim 25$.

\section{Distances} \label{sec:dist}

We derive the distance to Col~I using the prominent blue horizontal branch. We estimate the distance by performing a least-squares fit of our Col~I BHB stars (large diamonds in Figure~\ref{fig:coli_CMD}) to the empirical BHB ridgeline of the metal-poor ([Fe/H] = -2.34; \citealt{Carretta09}) globular cluster NGC~7078 (M~15) determined by \citet{bfs+14} using PanSTARRS photometry (assuming a distance modulus of $m-M=15.39$ to NGC~7078, here and throughout this work). We find a best-fitting distance modulus to Col~I of $m-M = 21.31\pm0.11$, 
corresponding to a distance of $d_{\rm Col~I} = 183\pm10$~kpc. In Figure~\ref{fig:coli_CMD}, we overplot the PanSTARRS ridgeline for the main sequence and RGB populations of NGC~7078, shifted to $m-M = 21.31$; the MSTO and RGB of Col~I very closely match the ridgelines.

To determine the distance to Tri~II, for which our data reach $>3$~mags below the MSTO, we perform a least-squares fit of the NGC~7078 ridgeline to all Tri~II candidates with $i_0 < 24$ (where the Tri~II main sequence is well separated from the unresolved background galaxy contamination, and below which the NGC~7078 ridgeline is unconstrained by PS1 data). This yields a distance modulus of $m-M = 17.27\pm0.11$, corresponding to $d_{\rm Tri~II} = 28.4\pm1.6$~kpc, which is similar to the distance of $30\pm2$~kpc estimated by \citet{lmi+15}. This corresponds to a Galactocentric distance of $d_{\rm GC, Tri~II} \approx 34.5$~kpc (assuming the Sun is 8 kpc from the Galactic center) at the Tri~II Galactic coordinates of $(l, b) = (140.9^\circ,-23.8^\circ)$. The NGC~7078 BHB ridgeline, shifted to $m-M = 17.27$, is overplotted in each panel of Figure~\ref{fig:triii_CMD}, with an old, metal-poor, alpha-enhanced Dartmouth isochrone shown in both panels to better illustrate the main sequence.

\begin{deluxetable*}{lrr}
\tablecaption{Properties of Col~I and Tri~II \label{tab:params}}
\tablehead{\colhead{Parameter} & \colhead{Col~I} & \colhead{Tri~II}}
\startdata
RA (hh:mm:ss) & $05:31:25.67 \pm 8.0''$ & $02:13:17.34 \pm 14.4''$ \\
Decl (dd:mm:ss) & $-28:02:33.1 \pm 11.4''$ & $+36:10:18.9 \pm 9.7''$ \\
$m-M$ (mag) & $21.31 \pm 0.11$ & $17.27 \pm 0.11$ \\
$d$ (kpc) & $183 \pm 10$ & $28.4 \pm 1.6$ \\
$M_{V}$ (mag) & $-4.2 \pm 0.2$ & $-1.2 \pm 0.4$\\
$r_{\rm h, exp}$ (arcmin) & $2.2 \pm 0.2$ & $2.5 \pm 0.3$ \\
$r_{\rm h, exp}$ (pc) & $117 \pm 17$ & $21 \pm 4$ \\
$\epsilon$ & $0.3 \pm 0.1$ & $0.3 \pm 0.1$  \\
$\theta$ (deg.) & $24^\circ \pm 9^\circ$ & $73^\circ \pm 17^\circ$ \\
$\langle\mu\rangle_{\rm eff, V}\tablenotemark{a}$ (mag arcsec$^{-2}$) & $29.0 \pm 0.6$ & $28.2 \pm 0.8$ \\
$r_{\rm c, King}$ (arcmin) & $2.1 \pm 0.6$ & $2.1 \pm 0.6$ \\
$r_{\rm c, King}$ (pc) & $112 \pm 40$ & $17 \pm 6$ \\
$r_{\rm t, King}$ (arcmin) & $8.9 \pm 2.5$ & $11.8 \pm 2.7$ \\
$r_{\rm t, King}$ (pc) & $472 \pm 160$ & $97 \pm 27$ \\
$M_{\rm HI} (M_\odot)$ & $<1.2\times10^4$ & $<3.1 \times 10^2$ \\
$M_{\rm HI}/L_{V} (M_\odot/L_\odot)$ & $<3.1$ & $<1.2$ 
\enddata
\tablenotetext{a}{Average surface brightness within the half-light radius.}
\end{deluxetable*}

\begin{figure*}[!t]
\begin{center}
\includegraphics[width=0.49\textwidth, trim=0.5cm {0.0cm} {0.5cm} {0.5cm}, clip]{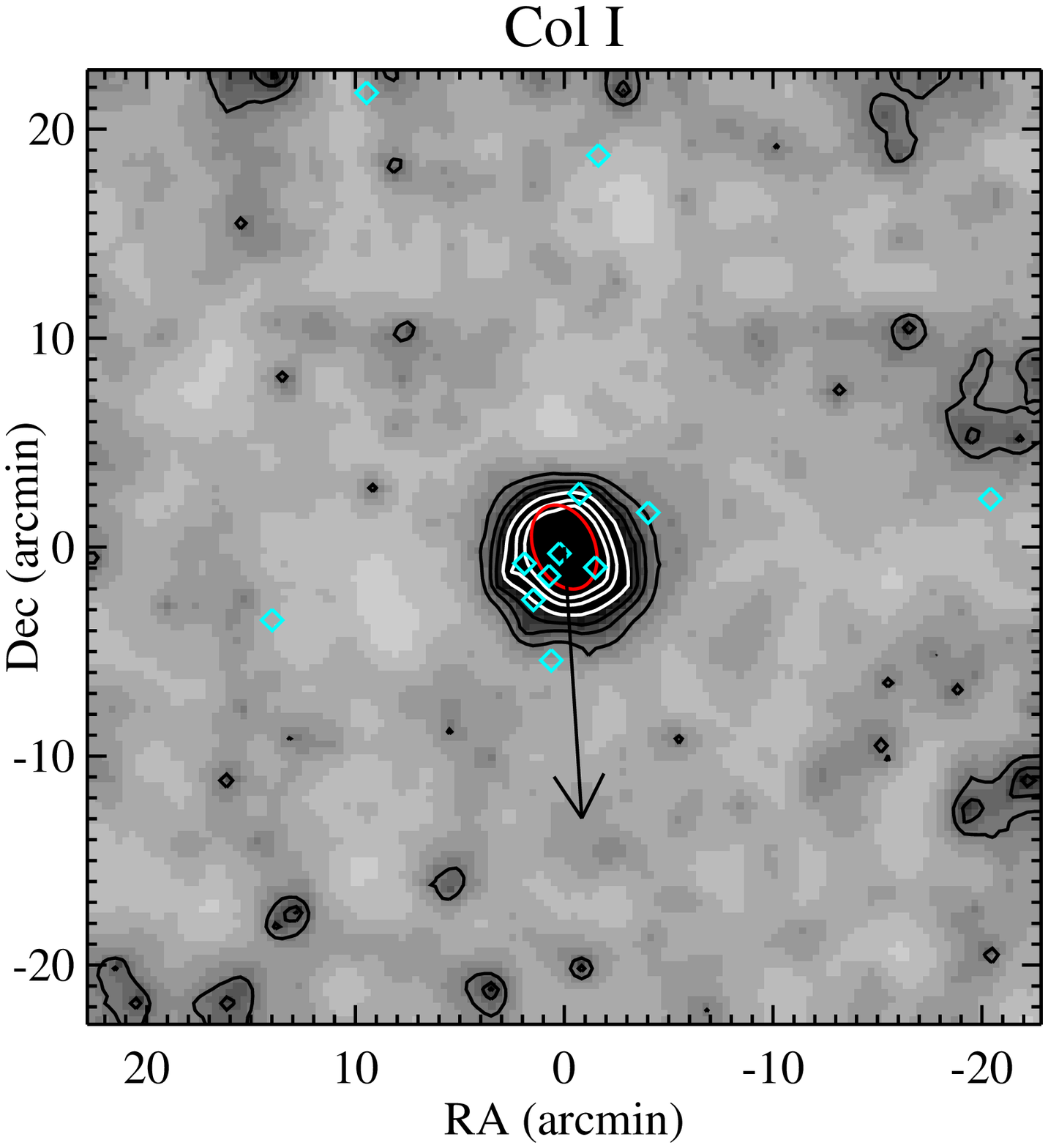}
\includegraphics[width=0.49\textwidth, trim=0.5cm {0.0cm} {0.5cm} {0.5cm}, clip]{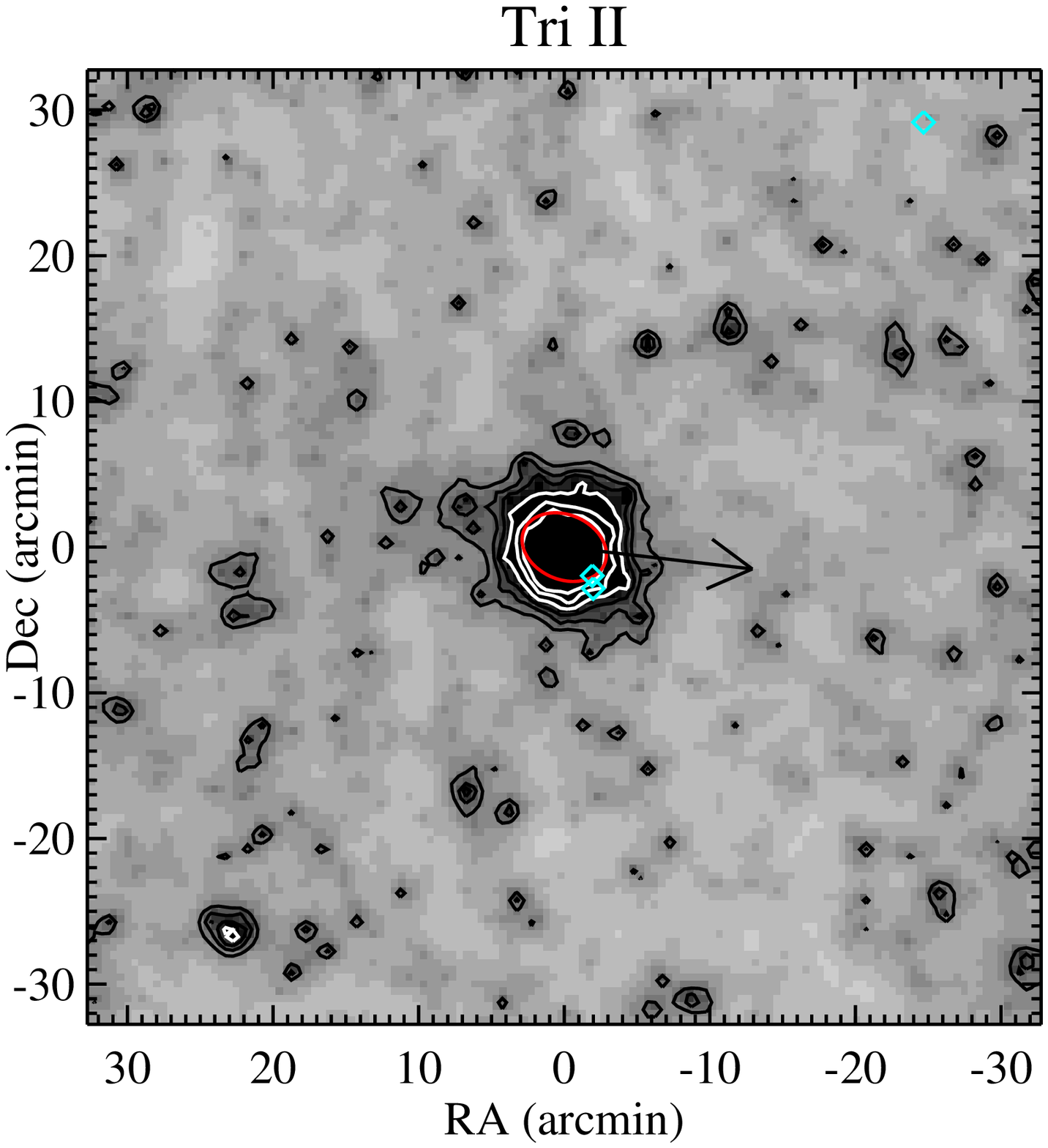}
\caption{Matched-filter density maps of point sources near Col~I and Tri~II, including sources brighter than $i_0 < 25.5$ for Col~I and $i_0 < 25$ for Tri~II. Bins for Col~I are $20''$ in RA and Dec, and for Tri~II, $30''$. Both maps have been smoothed with an exponential kernel with $45''$ scale length. Contours denote the $3\sigma, 5\sigma, 7\sigma, 10\sigma, 15\sigma$, and $20\sigma$ levels (though because the maps are smoothed, these should not be interpreted in the typical statistical sense). Each panel includes BHB candidates at the distance of the dwarf (i.e., stars within 0.15 mag of the NGC~7078 BHB ridgeline shifted to the appropriate distance, with $(g-i)_0 < 0$, and at $21.6 < i_0 < 22.7$ for Col~I, $17.5 < i_0 < 20.5$ for Tri~II) as cyan diamonds. In each panel, the red ellipse has semimajor axis equal to our measured half-light radius, with ellipticity and position angle as determined from the maximum likelihood analysis. The arrows in each panel point in the direction of the Galactic center.}
\label{fig:density}
\end{center}
\end{figure*}

\section{Structural Parameters and Luminosities} \label{sec:struct_params}

We estimate the structural parameters of both UFDs using the maximum likelihood method of \citet{mdr08} as implemented by \citet{Sand09} and \citet{Munoz10}. We include stars with $i_0 < 26.0$ that are within 12~arcmin of the center of Col~I, selected within a filter centered on the NGC~7078 ridgeline, that spans $\pm0.05$ mag at $i_0 = 18.0$, expanding linearly in width to $\pm0.15$ mag at $i_0 = 25.0$ (this ends up being 0.164 mag wide at $i_0 = 26$; the selection includes a total of 3242 stars, shown as gray points in Figure~\ref{fig:coli_CMD}). For Tri~II, we use a similar filter, but centered on a Dartmouth isochrone \citep{dotter08} with age 13~Gyr, [Fe/H] = -2.0, and $[\alpha$/Fe] = 0.4, with filter width increasing from $\pm0.05$ mag at $i_0 = 18$ to $\pm0.2$ mag at $i_0 = 25$. We include isochrone-filtered stars within $r < 15'$ of the Tri~II center with magnitudes $i_0 < 25$ (2550 input stars in total; gray points in Figure~\ref{fig:triii_CMD}). The resulting structural parameters are summarized in Table~\ref{tab:params}. These include the central position, half-light radius ($r_{\rm h}$), ellipticity ($\epsilon$), and position angle $\theta$ from fitting an exponential, and the King \citep{king62} model core and tidal radii ($r_{\rm c}$ and $r_{\rm t}$, respectively). Our deep, large-area data set satisfies all of the conditions (large field of view, total number of stars, and central-to-background density contrast) found by \citet{Munoz12} to be necessary for deriving accurate structural parameters.

Luminosities were derived using the technique of \citet{Sand09}. To do so, we generate synthetic stellar populations using IAC-Star \citep{AparicioGallart04_IACSTAR}\footnote{\url{http://iac-star.iac.es}}, sampling from a power-law IMF with slope -1.3 for $0.1~M_\odot < M < 0.5~M_\odot$ and -2.3 for $0.5~M_\odot < M < 120~M_\odot$ until we have 50,000 synthetic stars in our catalogs. For both UFDs, we use an old (13 Gyr), metal-poor ($Z = 0.0001$) stellar population. We then randomly sample the same number of stars as were included in our observed data sets (201 and 213 stars for Col~I and Tri~II, respectively) from these catalogs, within the magnitude ranges included in our parameter estimations. The derived luminosities represent the average and its standard deviation of the total flux from over 1000 samples for each dwarf. This accounts for the effects of CMD shot-noise. The resulting luminosities are tabulated in Table~\ref{tab:params}; we discuss the properties of each galaxy below.

\begin{figure*}[!t]
\begin{center}
\includegraphics[width=0.45\textwidth, trim=0.0cm {0.2cm} {0.0cm} {0.0cm}, clip]{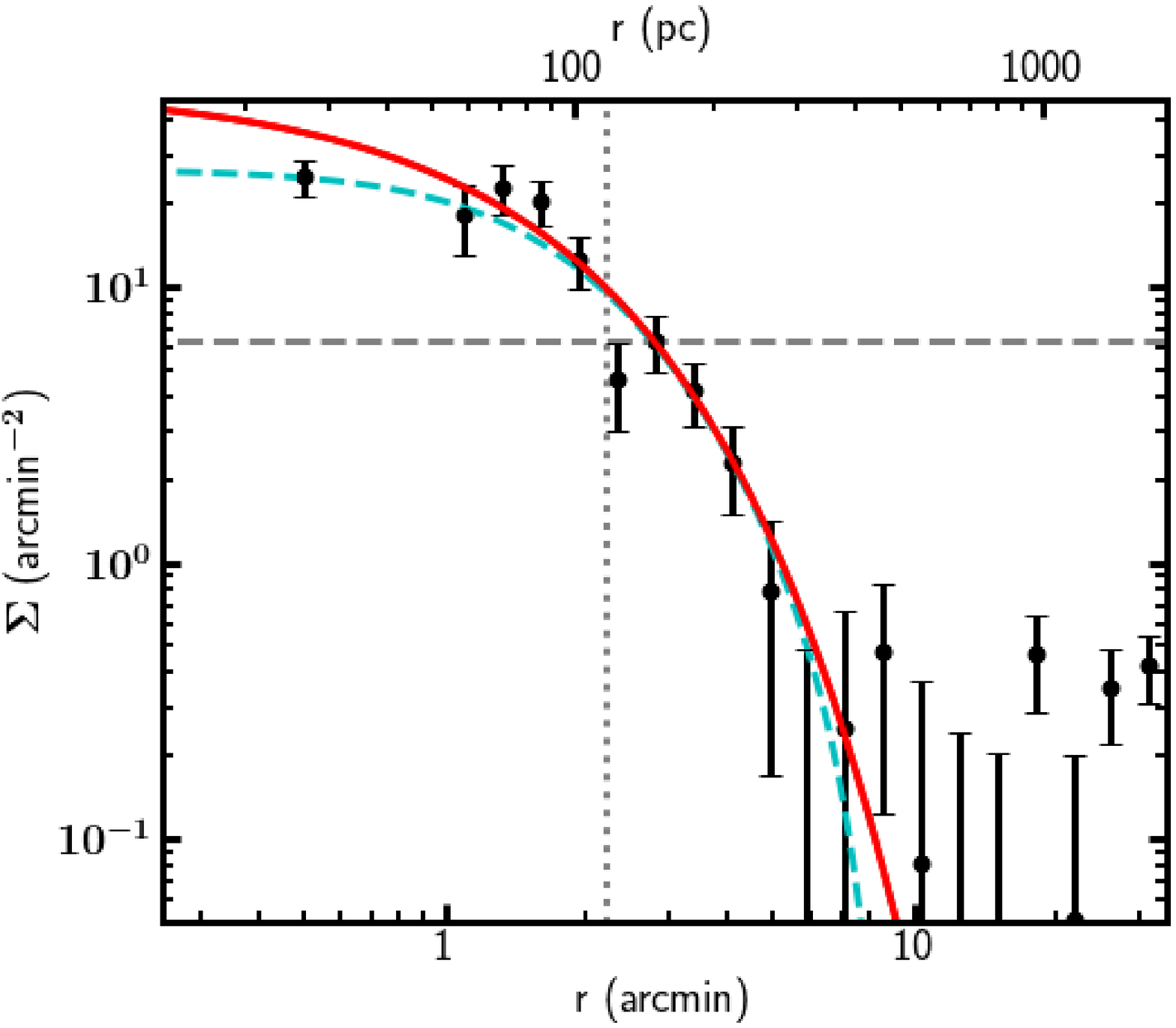}
\includegraphics[width=0.45\textwidth, trim=0.0cm {0.2cm} {0.0cm} {0.0cm}, clip]{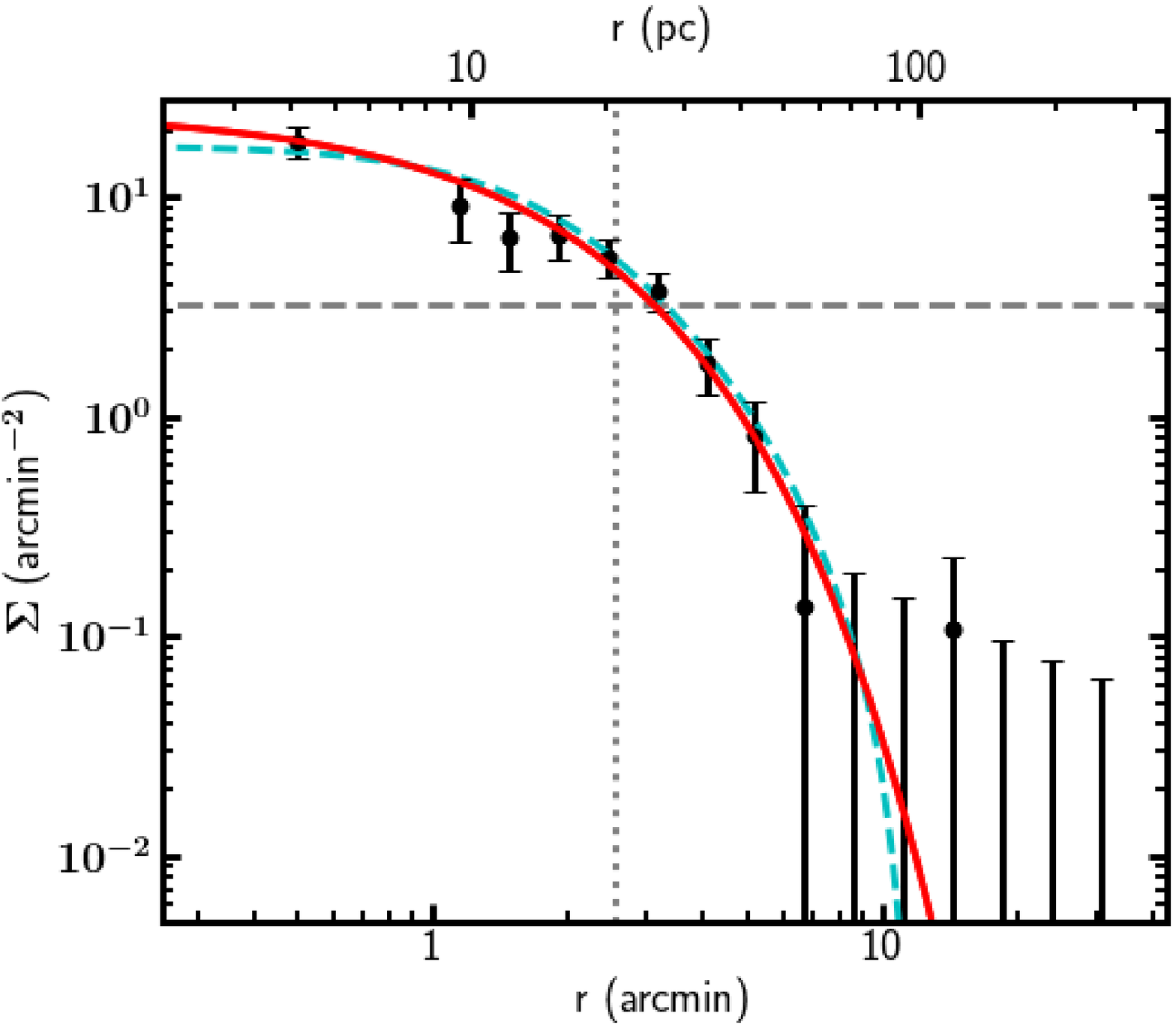}
\caption{{\it Left:} Surface density of Col~I stars between $18 < i_0 < 26$, using the isochrone filter described in the text. {\it Right:} Density profile of Tri~II stars with $18 < i_0 < 25$, also selected with an isochrone filter. In each panel, bins are elliptical (using the derived parameters in Table~\ref{tab:params}), centered on the dwarf. The solid red lines are exponential profiles using the half-light radii of our maximum likelihood fit, and the cyan dashed lines are the best-fit King models for each dwarf. The background as determined by the maximum likelihood method has been subtracted in each panel, allowing us to see structure well below the average background level. Each panel shows the subtracted background level as a horizontal dashed gray line, and the half-light radius as a vertical dotted gray line. Neither of the dwarfs shows obvious tidal disruption in the form of a ``break'' from the exponential profile at large radii, to more than an order of magnitude below the background surface density.
}
\label{fig:coli_triii_SBP}
\end{center}
\end{figure*}

\subsection{Col~I}

Our deep observations, which reach beyond the MSTO (at $i_0 \sim 25.5$), confirm that Col~I has a position in the size-luminosity plane (Fig.~\ref{fig:mv_rh}) consistent with being a distant, metal-poor ultra-faint dwarf. We find an extremely narrow RGB, and a prominent BHB (Fig.~\ref{fig:coli_CMD}). Unlike \citet{dbr+15}, we see no obvious evidence of red horizontal branch stars. In fact, Col~I appears to consist of a single, metal-poor stellar population, with no age or metallicity spread broadening its RGB. We find that the PS1 ridgeline of globular cluster NGC~7078 ([Fe/H] = -2.34; \citealt{Carretta09}) is an excellent match to Col~I, including its BHB, which implies that Col~I must be old and metal-poor as well. 
We derive a distance $d_{\rm Col~I} = 183\pm10$~kpc (see Section~\ref{sec:dist}), which agrees with the estimate of $182\pm18$~kpc from its discovery in DES \citep{dbr+15}. Likewise, our deeper, higher quality data yield a more precise measurement of the half-light radius: $r_{\rm h, Col~I} = 117\pm17$~pc which is consistent with the radius ($103\pm25$~pc) measured by \citet{dbr+15}. Col~I is rather round ($\epsilon \approx 0.3 \pm 0.1$), suggesting that this UFD shows no evidence of recent tidal stripping.
This is further confirmed in the density map (left panel of Fig.~\ref{fig:density}), which was created using the matched-filter technique of \citet{rog+02}; Col~I shows no irregularities in its density contours. The left panel of Figure~\ref{fig:coli_triii_SBP} shows the surface density of Col~I stars selected from the CMD as shown by the gray points in Figure~\ref{fig:coli_CMD}. The background as determined by the maximum likelihood method has been subtracted, and we overlay the exponential (red curve) and King (cyan dashed curve) model fits from the maximum likelihood analysis. Both of the model fits match the data well to surface densities at least an order of magnitude below the background level, with no obvious evidence of tidal disruption in the form of a break at large radii.

Our derived luminosity, $M_{\rm V} = -4.2\pm0.2$, is slightly fainter than the previous estimate ($M_{\rm V} = -4.5\pm0.2$; \citealt{dbr+15}), although the measurements are consistent within their uncertainties. Assuming a $V-$band absolute magnitude for the Sun of +4.83, our measurements imply a total $V$-band luminosity of $L_{\rm V} = 4.1^{+0.8}_{-0.7} \times 10^3~L_\odot$. We place Col~I in the context of other Local Group satellites in Figure~\ref{fig:mv_rh}, which shows the size--luminosity plane for all nearby dwarf galaxies and globular clusters. Col~I (large red star in Fig.~\ref{fig:mv_rh}) lands directly on the locus defined by other MW UFDs of similar luminosity.

\subsection{Tri~II}

Our Subaru/HSC observations probe at least 4 magnitudes deeper than Tri~II's MSTO. The precise photometry reveals a narrow main sequence consistent with an old, metal-poor stellar population (Figure~\ref{fig:triii_CMD}). A fit of the PS1 ridgeline for NGC~7078 provides a good match to the main sequence of Tri~II, yielding a distance estimate of $\sim28\pm2$~kpc. There is no clearly defined RGB locus in the CMD; indeed, it has been spectroscopically verified that most of the stars near the likely RGB location in the CMD are not members of Tri~II (see \citealt{kcs+15,mic+16,kcs+17}). There also seem to be few, if any, blue horizontal branch stars in Tri~II. As expected given the spectroscopic estimates of $\left<{\rm [Fe/H]}\right> \sim -2.5$ \citep{kcs+15,mic+16,kcs+17,vsm+17}, the ridgeline of the metal-poor ([Fe/H]$ \approx -2.3$) globular cluster NGC~7078 matches the data well, confirming that Tri~II contains a predominantly old, metal-poor stellar population. 

We also checked the mass function (MF) of stars in Tri~II. To do so, we fit the stellar mass as a function of $i$-band magnitude for a Dartmouth isochrone \citep{dotter08} with [Fe/H] = -2.3 and age 13.5 Gyr, shifted to our measured distance for Tri~II. Then, for each star within $3'$ of the Tri~II center, we assign a mass according to the mass-magnitude fit. We derive a mass function as the number of Tri~II stars in seven mass bins between $\sim$0.54 to 0.77 $M_\odot$ (magnitudes $20 < i_0 < 24$), subtracting off the average background MF from four equal-area background fields. We fit a power-law to the MF for Tri~II, and find $\alpha = 2.0\pm0.7$. Our Tri~II MF slope is consistent within the uncertainties with $\alpha_{\rm Her}=1.2^{+0.4}_{-0.5}$ and $\alpha_{\rm LeoIV}=1.3\pm0.8$ derived for the UFDs Hercules and Leo~IV by \citet{geha13}, but also consistent with a Salpeter IMF (i.e., $\alpha = 2.35$). Note, however, that we did not model the effect of unresolved binaries on the MF, which would steepen the slope toward a more bottom-heavy MF. There is no evidence for substantial preferential loss of low-mass stars, as would be expected for relaxed globular clusters in the last stages of tidal dissolution \citep[e.g.,][]{Contenta17}.

\begin{figure}[!t]
\begin{center}
\includegraphics[width=0.5\textwidth, trim=1.0cm {0.25cm} {0.0cm} {0.5cm}, clip]{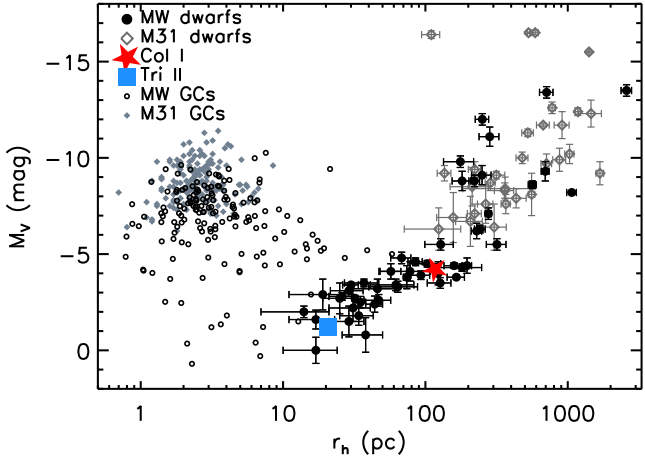}
\caption{Size-luminosity diagram placing the Tri~II (blue square) and Col~I (red star) properties in context with those of the MW dwarfs (filled black circles), MW GCs (small open black circles), M31 dwarfs (open gray diamonds), and M31 GCs (small filled gray diamonds). Tri~II and Col~I both sit along the sequence of UFDs found in the Local Group. Points for Col~I and Tri~II are the size of their error ranges. 
}
\label{fig:mv_rh}
\end{center}
\end{figure}

The half-light radius of Tri~II derived from our maximum likelihood analysis is $r_{\rm h} = 21\pm4$~pc, which is consistent with the original measurement from PS1 and LBT data ($r_{\rm h} = 34^{+9}_{-8}$~pc; \citealt{lmi+15}). We also find that Tri~II is fainter (though in agreement within the uncertainties) than was estimated by \citet{lmi+15}; we find $M_{\rm V} = -1.2\pm0.4$, while the previous estimate was $M_{\rm V} = -1.8\pm0.5$. This lower luminosity ($\sim260~L_\odot$; compare to $L_{\rm V} \sim 450~L_\odot$ for the Laevens et al. value) would seemingly make Tri~II an even greater outlier in the luminosity--metallicity relation for MW satellites, in which Tri~II may be offset by $\gtrsim0.5$~dex in mean metallicity from the UFD locus (e.g., Fig.~7 of \citealt{kcs+17}). However, given the paucity of stars with spectroscopic metallicities in Tri~II, the mean metallicity is rather dependent on the membership prospects of a small number of stars (or perhaps even a single star; e.g., \citealt{kcs+17}). 
As can be seen in Figure~\ref{fig:mv_rh}, a reduction in size along with a lower luminosity simply shifts Tri~II along the size-luminosity locus of Local Group dwarfs; Tri~II still lies squarely on the location populated by the lowest-luminosity MW UFDs. However, we note that when globular clusters are included in the size-luminosity diagram (as in Fig.~\ref{fig:mv_rh}), the faintest clusters overlap the region populated by the faintest UFDs. Thus, the position of Tri~II in this plane is perhaps suggestive that it is an UFD, but not a definitive indication.

One possible solution to the question of whether Tri~II is a tidally disrupting ultra-faint dwarf or a globular cluster could be found in the surface density distribution of Tri~II stars. In our matched-filter stellar density map of Tri~II (Fig.~\ref{fig:density}, right panel), there is no obvious tidal distortion evident. Our measured ellipticity (Table~\ref{tab:params}) of $\epsilon = 0.3\pm0.1$ already suggests that Tri~II is fairly round; the lack of extension 
in Fig.~\ref{fig:density} would seem to rule out strong/recent tidal disruption in this system. As another check on this scenario, we plot an azimuthally averaged, background-subtracted radial surface density profile in the right panel of Figure~\ref{fig:coli_triii_SBP}. This consists of the stars shown as gray points in the CMD of Figure~\ref{fig:triii_CMD}, binned in elliptical annuli with the measured ellipticity of Tri~II. The overplotted red line is an exponential profile with $r_{\rm h} = 2.5'$ as derived from our maximum likelihood analysis, with the cyan dashed curve representing the King model fit. Both model fits reproduce the density profile well to densities nearly two orders of magnitude below the background level. The lack of a ``break'' in the surface density profile may be further evidence that Tri~II has not recently suffered any tidal disruption.

\subsubsection{Tri~II's orbit and its dynamical status}

Its position and large radial velocity ($\left<v_{\rm GSR}\right> = -264$~km~s$^{-1}$; \citealt{kcs+17}) imply that Tri~II must be on a rather radial orbit, approaching its pericenter. The lack of obvious tidal debris is perhaps not surprising -- the ``break'' radius in the stellar density profile of a disrupting satellite drifts monotonically to larger radii after its minimum immediately after pericenter \citep[e.g.,][]{Lokas13}, such that any debris stripped on the previous pericentric passage have drifted far from the satellite by now. The narrow main sequence seen in Figure~\ref{fig:triii_CMD} suggests that the dwarf is also not extended along the line of sight. To confirm this, we measured the standard deviation about the NGC~7078 ridgeline for main sequence stars within $3'$ of Tri~II, and between $21 < i_0 < 24$. The scatter is $\sigma_{i_0} = 0.06$~mag, equal to roughly $\pm0.8$~kpc width about the mean distance. However, this does not account for the contribution of unresolved binaries to the main sequence width, nor the $\pm1.6$~kpc uncertainty in the distance itself. We thus conclude that we have not detected any line of sight extension of Tri~II.

\begin{figure*}[!t]
\begin{center}
\includegraphics[width=0.45\textwidth, trim=0.0cm {0.25cm} {0.0cm} {0.5cm}, clip]{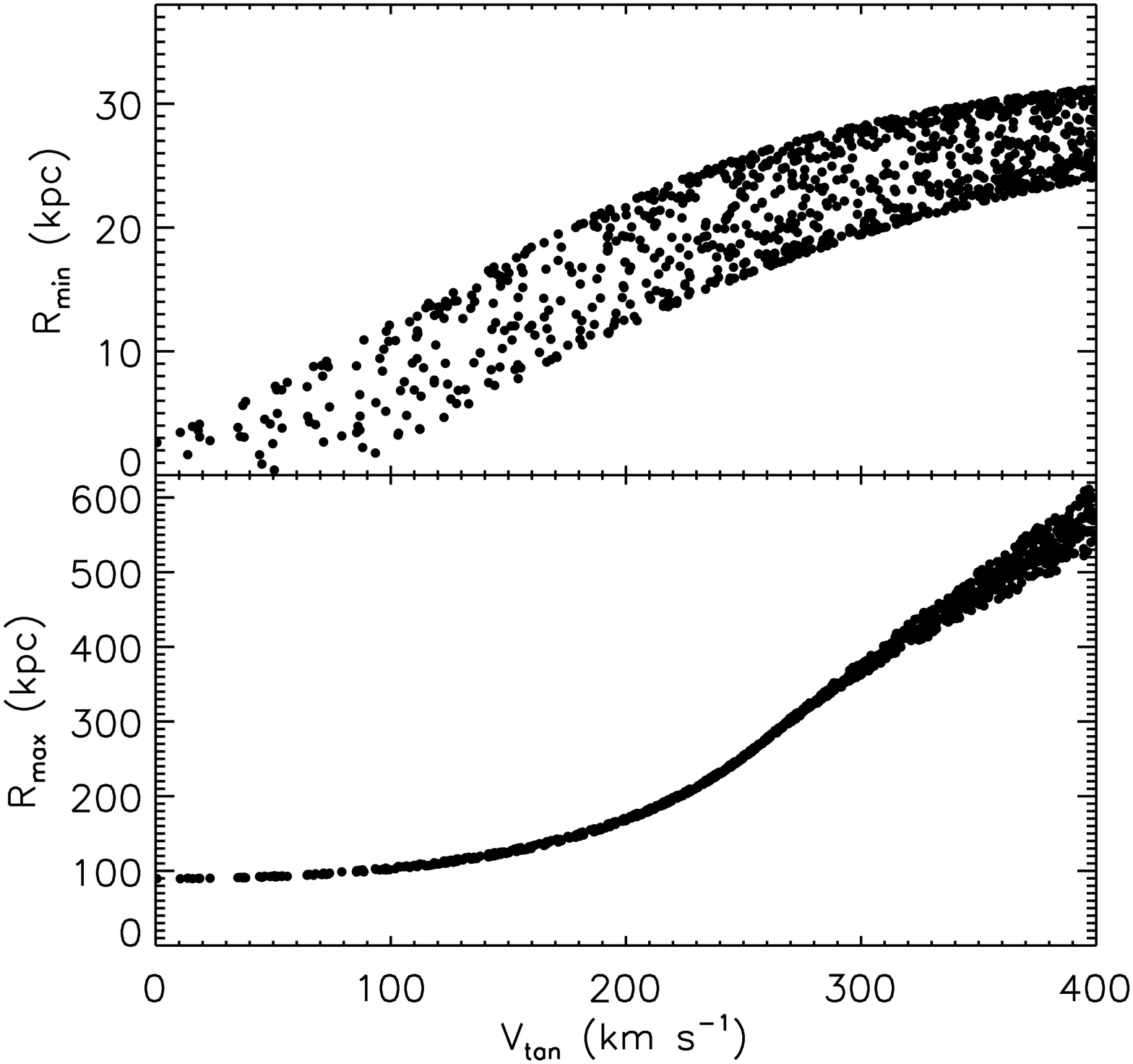}
\includegraphics[width=0.45\textwidth, trim=0.0cm {0.25cm} {0.0cm} {0.5cm}, clip]{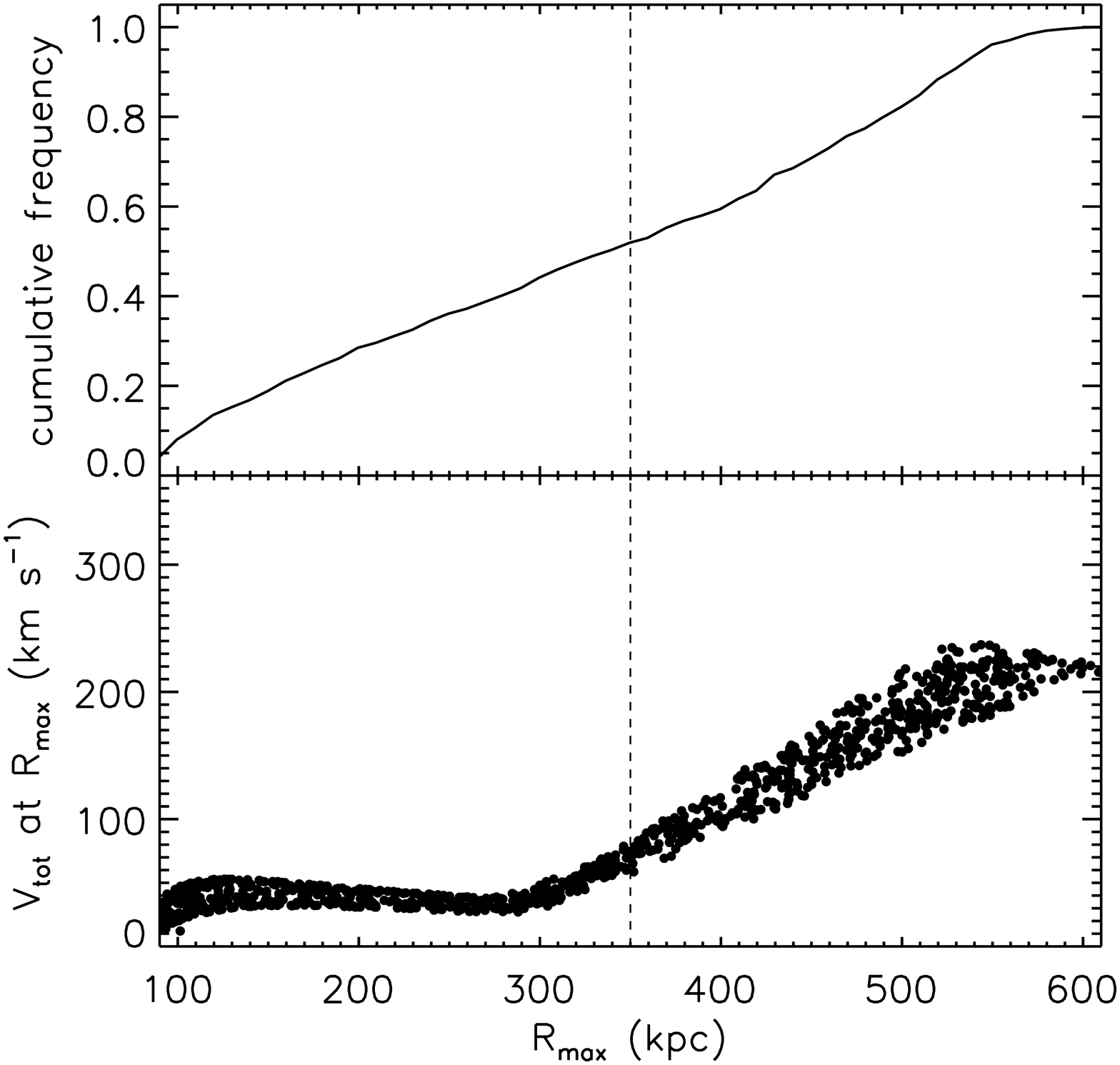}
\caption{Results from 1000 orbit simulations for Tri~II. Orbits were integrated for $\pm2$~Gyr, starting with the known $v_{\rm GSR}$, position, and distance of Tri~II, and random tangential velocities of $\left|V_{\rm tan}\right| < 400$~km~s$^{-1}$. {\it Left panels:} Minimum Galactocentric distance ($R_{\rm min}$; equivalent to the pericenter for a bound orbit) in the forward evolution of the orbit (upper panel), and maximum distance ($R_{\rm max}$; i.e., apocenter for bound orbits; lower panel) in the previous 2~Gyr. Many of the bound orbits approach the innermost regions of the Galaxy, while the minimum apocenter distance is $\sim95$~kpc. {\it Right panels:} Cumulative distribution of the maximum distance reached in the simulated orbits, and the total Galactocentric velocity at each $R_{\rm max}$ (which should be small for bound orbits). The vertical line at $R = 350$~kpc represents the approximate virial radius of the Milky Way (note that the velocities in the lower panel begin to rise near $R_{\rm max} \gtrsim 300$~kpc, suggesting that 350~kpc is a conservative estimate of the virial radius for our adopted potential). A total of 510 out of 1000 (or 51\%) of the orbits have $R_{\rm max} > 350$~kpc, meaning that more than half of the simulated Tri~II orbits are unbound (assuming a MW virial radius of 350 kpc).
}
\label{fig:orb_peri_apo}
\end{center}
\end{figure*}

We next consider what we can learn from the position and radial velocity of Tri~II about its orbit. To do so, we generate 1000 random tangential velocity vectors that are perpendicular to the direction of the velocity vector implied by $v_{\rm GSR}$, and have magnitude $\left|V_{\rm tan}\right| < 400$~km~s$^{-1}$ ($V_{\rm tan} = 400$~km~s$^{-1}$ would yield a total 3D Galactocentric velocity of 477~km~s$^{-1}$ for Tri~II, which exceeds the predicted Milky Way escape velocity at the distance of Tri~II; see, e.g., Fig.~9 of \citealt{Kafle14}). For each of the 1000 3D velocity vectors created by combining the radial velocity with $V_{\rm tan}$, we integrate an orbit starting from the position of Tri~II for $\pm3$~Gyr in the Galactic potential of \citet{Johnston98}\footnote{The gravitational potential implemented by \citet{Johnston98} includes a Miyamoto-Nagai disk \citep{miyamoto-nagai75}, a \citet{hernquist90} spheroid, and a logarithmic halo. Had we instead used the NFW halo potential of \citet{bovy15} or \citet{mcmillan17}, our results would be qualitatively similar, given that the difference between accelerations in the NFW and logarithmic halos is small in the outer regions of the Milky Way. Our arbitrary choice of potential was meant only to guide our intuition, and not to definitively ``fit'' the orbital behavior of Tri~II.}. Figure~\ref{fig:orb_peri_apo} (left panels) shows the minimum Galactocentric distance reached in the forward orbit integration (i.e., the pericenter for a bound orbit), and the maximum distance (apocenter if the satellite is bound) in the previous 2.0 Gyr, as a function of the total tangential velocity. As expected, the orbits are rather radial, with eccentricities between $0.75 < e < 0.99$. A number of the orbits pass very near the Galactic center. In the right panel of Figure~\ref{fig:orb_peri_apo} we show the cumulative distribution of the maximum distance, $R_{\rm max}$. More than half (51\%) of the orbits have $R_{\rm max} > 350$~kpc (i.e., greater than the approximate MW virial radius), suggesting that Tri~II could be on its first infall into the Milky Way's virial halo (though the good match of a King model to the surface density profile may suggest that the Galactic potential has imposed a truncation to the radial extent of Tri~II, which could argue against the first-infall scenario). We also note that $V_{\rm tan} \gtrsim 300$~km~s$^{-1}$ leads to unbound orbits; thus, we expect that if Tri~II is bound to the MW, it should have a total proper motion of $\mu_{\rm total} < \frac{V_{\rm tan}}{4.74 d}$, or $\mu_{\rm total} < 2.23$~mas~yr$^{-1}$ (assuming a distance of 28.4~kpc). The RGB of Tri~II may be within reach of $Gaia$, for which expected proper motion uncertainties are $\lesssim0.3$~mas per star at $G=20$ \citep{Perryman01},\footnote{\url{https://www.cosmos.esa.int/web/gaia/science-performance}} while LSST will achieve $\lesssim0.5$~mas per star accuracy at $r\sim23$ \citep[e.g.,][]{LSSTScienceBook}, well below the turnoff of Tri~II.

\begin{figure}[!t]
\begin{center}
\includegraphics[width=0.45\textwidth, trim=0.0cm {0.25cm} {0.0cm} {0.5cm}, clip]{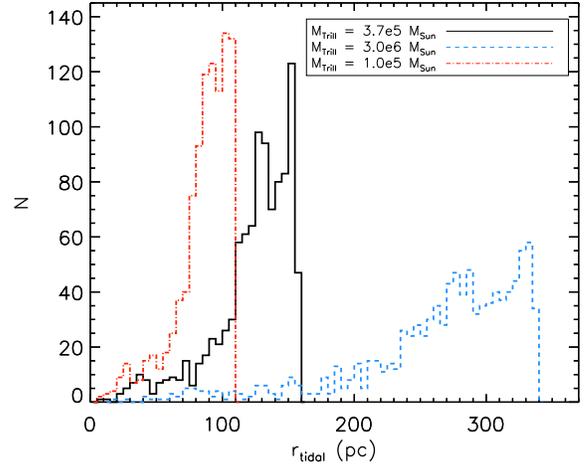}
\caption{Distribution of tidal radii at pericenter for the 1000 simulated orbits of Tri~II, assuming three different masses for the dwarf. The tidal radius is smaller than the observed extent of Tri~II (e.g., Fig.~\ref{fig:coli_triii_SBP}) and our measured King tidal radius of $r_{\rm t} = 97\pm27$~pc only for the lowest mass progenitor (and a small fraction of the intermediate mass simulated UFDs). 
}
\label{fig:rtidal_hist}
\end{center}
\end{figure}

With these simulated orbits in hand, we can assess the possible tidal interaction of Tri~II within the Galactic potential. In Figure~\ref{fig:rtidal_hist} we show the distribution of tidal radii of Tri~II at pericenter for the 1000 simulated orbits. Tidal radii (for a logarithmic Galactic potential) are calculated using the formula of \citet{OhLinAarseth95}:

\begin{equation}
r_{\rm tidal} = a \bigg[\frac{M_{\rm sat}}{M_{\rm Gal}}\bigg]^{1/3} \left\{\frac{(1-e)^2}{[(1+e)^2/2e]\ln{[(1+e)/(1-e)]}+1}\right\}^{1/3},
\end{equation}

\noindent where $a$ and $e$ are the orbital semimajor axis and eccentricity, respectively, $M_{\rm sat}$ is the satellite's total mass, and $M_{\rm Gal}$ is the mass of the Milky Way within the semimajor axis. We calculate $M_{\rm Gal}$ using 

\begin{equation}
M_{\rm Gal} (r) \approx 1.1\times10^{10} \bigg( \frac{r}{1~{\rm kpc}} \bigg) M_\odot,
\end{equation}

\noindent from \citet{Burkert97}, assuming $v_{\rm circ, MW} = 220$~km~s$^{-1}$. Because the mass of Tri~II is uncertain, we choose three values:\footnote{Note that these literature values, which are based on measured velocity dispersions using, e.g., the method of \citet{wolf10}, correspond to the mass within the half-light radius. Thus the total mass of Tri~II, if it has a significant dark matter halo, is likely much larger, which would make our estimates of the tidal radii lower limits.} (i) $M_{\rm TriII} = 3.7\times10^5~M_\odot$ (black histogram), corresponding to the upper limit from \citet{kcs+17}; (ii) $M_{\rm TriII} = 3.0\times10^6~M_\odot$ (\citealt{mic+16}; blue dashed histogram\footnote{Even though this mass estimate has been superseded by that of \citet{kcs+17}, we have included it for completeness, as the true velocity dispersion of Tri~II is still unknown.}); and a low mass (iii) $M_{\rm TriII} = 1.0\times10^5~M_\odot$ (red, dot-dashed histogram). Recall that the stellar density in Figure~\ref{fig:coli_triii_SBP} shows no break from an exponential profile out to $\sim90$~pc, where it blends with the background density. Furthermore, our King model maximum likelihood fits to the stellar density profile yield a tidal radius of $97\pm27$~pc for Tri~II. If the mass of Tri~II is as low as $10^5~M_\odot$, Fig.~\ref{fig:rtidal_hist} suggests that its tidal radius would be less than 90~pc for nearly all simulated orbits, in which case we might expect to see tidal debris (assuming the surface brightness of the debris was within reach of our observations).\footnote{Note that assuming a $V$-band stellar mass-to-light ratio of $2~M_\odot/L_{\rm V, \odot}$, our measured luminosity of Tri~II ($L_{\rm V} \sim 260~L_\odot$) corresponds to a stellar mass of $M_{\rm *, Tri~II} \sim 520~M_\odot$. A dark matter-free satellite with this stellar mass would have tidal radii 5.8 times smaller than those of the $1\times10^5~M_\odot$ satellite shown in Figure~\ref{fig:rtidal_hist}, and would thus be highly susceptible to tidal disruption.} For the more massive simulated satellites, few of the tidal radii are smaller than 90~pc, suggesting that even if Tri~II is on a bound orbit, it is unlikely to suffer significant tidal disruption. Understanding the puzzling Tri~II system thus depends critically on resolving its stellar velocity dispersion to accurately derive its mass.

\section{H\textsc{I} upper limits} \label{sec:HI}

We searched for atomic gas reservoirs in Col~I and Tri~II in publicly available data from the Southern hemisphere Galactic All Sky Survey (GASS; \citealt{mcclure09}) and Northern hemisphere Effelsberg Bonn H\textsc{I} Survey (EBHIS; \citealt{winkel16}), respectively. We find no H\textsc{I} emission peaks that are statistically significant at the 5$\sigma$ level along the line of sight to either dwarf when the data are smoothed to a spectral resolution of 15$\,\mathrm{km\,s^{-1}}$. Adopting the distances to Col~I and Tri~II determined in \S\ref{sec:dist}, the corresponding 5$\sigma$, single-channel upper limits on the HI mass $M_{\rm HI}$ and on the relative gas content $M_{\rm HI}/L_{\rm V}$ are given in Table~\ref{tab:params}.

We note that we find an unresolved emission peak at lower statistical significance ($3.5\sigma$) along the line-of-sight to Tri~II at $V_{\rm helio} \sim -395~\mathrm{km~s^{-1}}$ in the smoothed EBHIS data. This velocity differs by only $\sim 15~\mathrm{km~s^{-1}}$ from the systemic velocity of the dwarf measured from stellar kinematics \citep{kcs+17}. Given the presence of similar peaks across the $20\,{\rm deg}^{2}$ EBHIS datacube containing the Tri~II line-of-sight as well as the detection of high-velocity Galactic H\textsc{I} features and gas associated with M31 at similar velocities in this region \citep{WvW97,kerp16}, we conclude that this peak is unlikely to stem from a gas reservoir in Tri~II.

While these H\textsc{I} upper limits do not place as strong constraints on the neutral gas content as those derived by \citet{Spekkens14} for other MW dwarfs, they are consistent with the overall lack of H\textsc{I} in {\it all} dwarf spheroidals observed within the MW virial radius (see Fig.~2 of \citealt{Spekkens14}). Given the lack of obvious young stellar populations in either Col~I or Tri~II, it is likely that the two systems do not contain significant reservoirs of gas.

\section{Discussion and Conclusions} \label{sec:conclusions}

We present deep Subaru/Hyper Suprime-Cam imaging of Milky Way companions Col~I and Tri~II, from which we derive the structural parameters (summarized in Table~\ref{tab:params}) and map the stellar density fields around these two satellites (Fig.~\ref{fig:density}). Our deep photometry reaching beyond the MSTO of Col~I, at a distance of $d_{\rm Col~I} = 183\pm10$~kpc, shows this satellite to have properties consistent with most Galactic UFDs. It is apparently made up of an old, metal-poor stellar population, including a prominent BHB (see Fig.~\ref{fig:coli_CMD}), which we use to estimate the distance to Col~I. We find $M_{\rm V} = -4.2\pm0.2$ for Col~I, and a half-light radius of $r_{\rm h} = 117\pm17$~pc, placing it directly on the observed size-luminosity relation for Local Group dwarfs (Fig.~\ref{fig:mv_rh}). The stellar density map of Col~I shows no evidence of obvious distortions or tidal disruption. We also search archival data for evidence of neutral hydrogen in Col~I, and derive an upper limit of $M_{\rm H\sc I} < 1.2\times10^4~M_\odot$, and $M_{\rm H\sc I}/L_{\rm V} < 3.1~M_\odot/L_\odot$. Overall, Col~I appears to be typical of old, metal-poor, gas-free UFDs in the Milky Way, but currently resides at a large distance from the Galaxy.

We derive a distance to Tri~II of $d_{\rm Tri~II} = 28.4\pm1.6$~kpc via least-squares fitting of $>3$~mags of the resolved main sequence. Tri~II's stellar population is well-matched by the empirical ridgeline of old, metal-poor globular cluster NGC~7078 (from \citealt{bfs+14}), as expected from its measured spectroscopic metallicity of $\left<{\rm [Fe/H]}\right> = -2.24$ \citep{kcs+17}. 
Tri~II is extremely faint, with $M_{\rm V} = -1.2\pm0.4$ (or $L_{\rm V} \sim 260~L_\odot$), and compact ($r_{\rm h} = 21\pm4$~pc). The stellar density map of Tri~II is of particular interest, because one possible solution to its higher metallicity relative to the Local Group relation at its measured $M_{\rm V}$ (e.g., \citealt{kcs+17}) is that Tri~II may have been more massive/luminous in the past, but suffered tidal mass loss. We see no evidence of tidal debris in either the stellar density map (Fig.~\ref{fig:density}) or radial stellar density profile (Fig.~\ref{fig:coli_triii_SBP}). 
We search archival H\textsc{I} observations near Tri~II, and find upper limits of $M_{\rm H\sc I} < 3.1\times10^2~M_\odot$ and $M_{\rm H\sc I}/L_{\rm V} < 1.2~M_\odot/L_\odot$. Both this and the Col~I upper limit are consistent with the lack of observed H\textsc{I} in all dSphs within the MW virial radius \citep{Spekkens14}, though not particularly stringent limits on their neutral gas content.

We further explore the dynamical state of Tri~II via a suite of orbital simulations based on its position and radial velocity, for 1000 different values of its tangential velocity. We find that more than half of the simulated orbits place Tri~II on its first infall into the MW potential. If so, the atypical properties of Tri~II relative to other UFDs may be due to different environmental effects. The surface density profile of Tri~II is well-matched by an exponential profile, with no evidence for tidal debris in the form of a break in the profile at large radii. We show that the predicted tidal radii from our simulated orbits are larger than the observed extent of Tri~II for all but the lowest mass satellites. We additionally find that Tri~II has a present-day stellar mass function similar to those of other UFDs, in contrast to the flatter mass functions typical of globular clusters (which arise due to dynamical evolution, and, in some cases, preferential loss of low-mass stars to tidal stripping). Taken together, the evidence we have presented in this work suggests that Tri~II is a dwarf galaxy with no evidence of being affected by tides. 

In this contribution, we have presented deep Subaru/Hyper Suprime-Cam observations of Milky Way satellites Columba~I and Triangulum~II. Col~I has properties typical of MW ultra-faint dwarfs, and Tri~II has properties more like a dwarf galaxy than a globular cluster. Our work highlights the precision that can be attained in measurements of UFD structural parameters with high quality, deep photometry reaching $\gtrsim 2$~mags deeper than previous data sets. In addition, with the large field of view covered by our HSC imaging, we find no evidence for the presence of significant tidal debris within several tidal radii of each of these dwarfs (within our surface brightness limits).

\acknowledgments

We thank the referee for a careful reading of the manuscript, and comments that helped us improve the paper. We thank Fumiaki Nakata and Rita Morris for assistance at the Subaru Telescope. JLC and BW are partially supported by NSF Faculty Early Career Development (CAREER) award AST-1151462.
DJS acknowledges support from NSF grant AST-1412504. JS acknowledges support from NSF grant AST-1514763 and a Packard Fellowship. AJR was supported by NSF grant AST-1616710 and as a Research Corporation for Science Advancement Cottrell Scholar.  

We thank Edouard Bernard for kindly sharing the PS1 globular cluster fiducials, and Keith Bechtol and David Nidever for helpful discussions. The authors wish to recognize and acknowledge the very significant cultural role and reverence that the summit of Mauna Kea has always had within the indigenous Hawaiian community.  We are most fortunate to have the opportunity to conduct observations from this mountain. 

This research has made use of NASA's Astrophysics Data System, and \textit{Astropy}, a community-developed core Python package for Astronomy \citep{Astropy}. This work has made use of the IAC-STAR Synthetic CMD computation code. IAC-STAR is supported and maintained by the computer division of the Instituto de Astrof\'isica de Canarias. 

Some of the data presented in this paper were obtained from the Mikulski Archive for Space Telescopes (MAST). STScI is operated by the Association of Universities for Research in Astronomy, Inc., under NASA contract NAS5-26555. Support for MAST for non-HST data is provided by the NASA Office of Space Science via grant NNX09AF08G and by other grants and contracts. 

The Pan-STARRS1 Surveys have been made possible through contributions of the Institute for Astronomy, the University of Hawaii, the Pan-STARRS Project Office, the Max-Planck Society and its participating institutes, the Max Planck Institute for Astronomy, Heidelberg and the Max Planck Institute for Extraterrestrial Physics, Garching, The Johns Hopkins University, Durham University, the University of Edinburgh, Queen's University Belfast, the Harvard-Smithsonian Center for Astrophysics, the Las Cumbres Observatory Global Telescope Network Incorporated, the National Central University of Taiwan, the Space Telescope Science Institute, the National Aeronautics and Space Administration under Grant No. NNX08AR22G issued through the Planetary Science Division of the NASA Science Mission Directorate, the National Science Foundation under Grant AST-1238877, the University of Maryland, Eotvos Lorand University (ELTE), and the Los Alamos National Laboratory.



\vspace{5mm}
\facilities{Subaru (Hyper Suprime-Cam)}

\software{Astropy \citep{Astropy}, IDL astronomy users library \citep{IDLastro}, iPython \citep{Perez07_iPython}, Matplotlib \citep{Matplotlib}, NumPy \citep{vanderWalt11_numpy}, Topcat \citep{Topcat}.}

\bibliographystyle{aasjournal}

\begin{thebibliography}{}
\expandafter\ifx\csname natexlab\endcsname\relax\def\natexlab#1{#1}\fi
\providecommand{\url}[1]{\href{#1}{#1}}

\bibitem[{{Abazajian} {et~al.}(2009){Abazajian}, {Adelman-McCarthy},
  {Ag{\"u}eros}, {Allam}, {Allende Prieto}, {An}, {Anderson}, {Anderson},
  {Annis}, {Bahcall}, \& et~al.}]{SDSSII}
{Abazajian}, K.~N., {Adelman-McCarthy}, J.~K., {Ag{\"u}eros}, M.~A., {et~al.}
  2009, \apjs, 182, 543

\bibitem[{{Aihara} {et~al.}(2017){Aihara}, {Armstrong}, {Bickerton}, {Bosch},
  {Coupon}, {Furusawa}, {Hayashi}, {Ikeda}, {Kamata}, {Karoji}, {Kawanomoto},
  {Koike}, {Komiyama}, {Lupton}, {Mineo}, {Miyatake}, {Miyazaki}, {Morokuma},
  {Obuchi}, {Oishi}, {Okura}, {Price}, {Takata}, {Tanaka}, {Tanaka}, {Tanaka},
  {Uchida}, {Uraguchi}, {Utsumi}, {Wang}, {Yamada}, {Yamanoi}, {Yasuda},
  {Arimoto}, {Chiba}, {Finet}, {Fujimori}, {Fujimoto}, {Furusawa}, {Goto},
  {Goulding}, {Gunn}, {Harikane}, {Hattori}, {Hayashi}, {Helminiak}, {Higuchi},
  {Hikage}, {Ho}, {Hsieh}, {Huang}, {Huang}, {Imanishi}, {Iwata}, {Jaelani},
  {Jian}, {Kashikawa}, {Katayama}, {Kojima}, {Konno}, {Koshida}, {Leauthaud},
  {Lee}, {Lin}, {Lin}, {Mandelbaum}, {Matsuoka}, {Medezinski}, {Miyama},
  {Momose}, {More}, {More}, {Mukae}, {Murata}, {Murayama}, {Nagao}, {Nakata},
  {Niikura}, {Nishizawa}, {Oguri}, {Okabe}, {Ono}, {Onodera}, {Onoue}, {Ouchi},
  {Pyo}, {Shibuya}, {Shimasaku}, {Simet}, {Speagle}, {Spergel}, {Strauss},
  {Sugahara}, {Sugiyama}, {Suto}, {Suzuki}, {Tait}, {Takada}, {Terai}, {Toba},
  {Turner}, {Uchiyama}, {Umetsu}, {Urata}, {Usuda}, {Yeh}, \&
  {Yuma}}]{Aihara17}
{Aihara}, H., {Armstrong}, R., {Bickerton}, S., {et~al.} 2017, ArXiv e-prints,
  arXiv:1702.08449

\bibitem[{{Alam} {et~al.}(2015){Alam}, {Albareti}, {Allende Prieto}, {Anders},
  {Anderson}, {Anderton}, {Andrews}, {Armengaud}, {Aubourg}, {Bailey}, \&
  et~al.}]{SDSSIII}
{Alam}, S., {Albareti}, F.~D., {Allende Prieto}, C., {et~al.} 2015, \apjs, 219,
  12

\bibitem[{{Albert} {et~al.}(2017){Albert}, {Anderson}, {Bechtol},
  {Drlica-Wagner}, {Meyer}, {S{\'a}nchez-Conde}, {Strigari}, {Wood}, {Abbott},
  {Abdalla}, {Benoit-L{\'e}vy}, {Bernstein}, {Bernstein}, {Bertin}, {Brooks},
  {Burke}, {Carnero Rosell}, {Carrasco Kind}, {Carretero}, {Crocce}, {Cunha},
  {D'Andrea}, {da Costa}, {Desai}, {Diehl}, {Dietrich}, {Doel}, {Eifler},
  {Evrard}, {Fausti Neto}, {Finley}, {Flaugher}, {Fosalba}, {Frieman},
  {Gerdes}, {Goldstein}, {Gruen}, {Gruendl}, {Honscheid}, {James}, {Kent},
  {Kuehn}, {Kuropatkin}, {Lahav}, {Li}, {Maia}, {March}, {Marshall}, {Martini},
  {Miller}, {Miquel}, {Neilsen}, {Nord}, {Ogando}, {Plazas}, {Reil}, {Romer},
  {Rykoff}, {Sanchez}, {Santiago}, {Schubnell}, {Sevilla-Noarbe}, {Smith},
  {Soares-Santos}, {Sobreira}, {Suchyta}, {Swanson}, {Tarle}, {Vikram},
  {Walker}, {Wechsler}, {Fermi-LAT Collaboration}, \& {DES
  Collaboration}}]{Albert17}
{Albert}, A., {Anderson}, B., {Bechtol}, K., {et~al.} 2017, \apj, 834, 110

\bibitem[{{Amorisco}(2017)}]{Amorisco17}
{Amorisco}, N.~C. 2017, \apj, 844, 64

\bibitem[{{Aparicio} \& {Gallart}(2004)}]{AparicioGallart04_IACSTAR}
{Aparicio}, A., \& {Gallart}, C. 2004, \aj, 128, 1465

\bibitem[{{Astropy Collaboration} {et~al.}(2013){Astropy Collaboration},
  {Robitaille}, {Tollerud}, {Greenfield}, {Droettboom}, {Bray}, {Aldcroft},
  {Davis}, {Ginsburg}, {Price-Whelan}, {Kerzendorf}, {Conley}, {Crighton},
  {Barbary}, {Muna}, {Ferguson}, {Grollier}, {Parikh}, {Nair}, {Unther},
  {Deil}, {Woillez}, {Conseil}, {Kramer}, {Turner}, {Singer}, {Fox}, {Weaver},
  {Zabalza}, {Edwards}, {Azalee Bostroem}, {Burke}, {Casey}, {Crawford},
  {Dencheva}, {Ely}, {Jenness}, {Labrie}, {Lim}, {Pierfederici}, {Pontzen},
  {Ptak}, {Refsdal}, {Servillat}, \& {Streicher}}]{Astropy}
{Astropy Collaboration}, {Robitaille}, T.~P., {Tollerud}, E.~J., {et~al.} 2013,
  \aap, 558, A33

\bibitem[{{Beasley} {et~al.}(2016){Beasley}, {Romanowsky}, {Pota}, {Navarro},
  {Martinez Delgado}, {Neyer}, \& {Deich}}]{beasley16}
{Beasley}, M.~A., {Romanowsky}, A.~J., {Pota}, V., {et~al.} 2016, \apjl, 819,
  L20

\bibitem[{{Bechtol} {et~al.}(2015){Bechtol}, {Drlica-Wagner}, {Balbinot},
  {Pieres}, {Simon}, {Yanny}, {Santiago}, {Wechsler}, {Frieman}, {Walker},
  {Williams}, {Rozo}, {Rykoff}, {Queiroz}, {Luque}, {Benoit-L{\'e}vy},
  {Tucker}, {Sevilla}, {Gruendl}, {da Costa}, {Fausti Neto}, {Maia}, {Abbott},
  {Allam}, {Armstrong}, {Bauer}, {Bernstein}, {Bernstein}, {Bertin}, {Brooks},
  {Buckley-Geer}, {Burke}, {Carnero Rosell}, {Castander}, {Covarrubias},
  {Drsquo Andrea}, {DePoy}, {Desai}, {Diehl}, {Eifler}, {Estrada}, {Evrard},
  {Fernandez}, {Finley}, {Flaugher}, {Gaztanaga}, {Gerdes}, {Girardi},
  {Gladders}, {Gruen}, {Gutierrez}, {Hao}, {Honscheid}, {Jain}, {James},
  {Kent}, {Kron}, {Kuehn}, {Kuropatkin}, {Lahav}, {Li}, {Lin}, {Makler},
  {March}, {Marshall}, {Martini}, {Merritt}, {Miller}, {Miquel}, {Mohr},
  {Neilsen}, {Nichol}, {Nord}, {Ogando}, {Peoples}, {Petravick}, {Plazas},
  {Romer}, {Roodman}, {Sako}, {Sanchez}, {Scarpine}, {Schubnell}, {Smith},
  {Soares-Santos}, {Sobreira}, {Suchyta}, {Swanson}, {Tarle}, {Thaler},
  {Thomas}, {Wester}, {Zuntz}, \& {The DES Collaboration}}]{Bechtol15}
{Bechtol}, K., {Drlica-Wagner}, A., {Balbinot}, E., {et~al.} 2015, \apj, 807,
  50

\bibitem[{{Belokurov} {et~al.}(2006){Belokurov}, {Zucker}, {Evans},
  {Wilkinson}, {Irwin}, {Hodgkin}, {Bramich}, {Irwin}, {Gilmore}, {Willman},
  {Vidrih}, {Newberg}, {Wyse}, {Fellhauer}, {Hewett}, {Cole}, {Bell}, {Beers},
  {Rockosi}, {Yanny}, {Grebel}, {Schneider}, {Lupton}, {Barentine},
  {Brewington}, {Brinkmann}, {Harvanek}, {Kleinman}, {Krzesinski}, {Long},
  {Nitta}, {Smith}, \& {Snedden}}]{Belokurov06_Bootes}
{Belokurov}, V., {Zucker}, D.~B., {Evans}, N.~W., {et~al.} 2006, \apjl, 647,
  L111

\bibitem[{{Belokurov} {et~al.}(2007){Belokurov}, {Zucker}, {Evans}, {Kleyna},
  {Koposov}, {Hodgkin}, {Irwin}, {Gilmore}, {Wilkinson}, {Fellhauer},
  {Bramich}, {Hewett}, {Vidrih}, {De Jong}, {Smith}, {Rix}, {Bell}, {Wyse},
  {Newberg}, {Mayeur}, {Yanny}, {Rockosi}, {Gnedin}, {Schneider}, {Beers},
  {Barentine}, {Brewington}, {Brinkmann}, {Harvanek}, {Kleinman}, {Krzesinski},
  {Long}, {Nitta}, \& {Snedden}}]{Belokurov07_5MWsats}
---. 2007, \apj, 654, 897

\bibitem[{{Bernard} {et~al.}(2014){Bernard}, {Ferguson}, {Schlafly}, {Platais},
  {Bell}, {Martin}, {Rix}, {Slater}, {Burgett}, {Chambers}, {Draper}, {Hodapp},
  {Kaiser}, {Kudritzki}, {Magnier}, {Metcalfe}, {Tonry}, {Wainscoat}, \&
  {Waters}}]{bfs+14}
{Bernard}, E.~J., {Ferguson}, A.~M.~N., {Schlafly}, E.~F., {et~al.} 2014,
  \mnras, 442, 2999

\bibitem[{{Bosch} {et~al.}(2017){Bosch}, {Armstrong}, {Bickerton}, {Furusawa},
  {Ikeda}, {Koike}, {Lupton}, {Mineo}, {Price}, {Takata}, {Tanaka}, {Yasuda},
  {AlSayyad}, {Becker}, {Coulton}, {Coupon}, {Garmilla}, {Huang}, {Krughoff},
  {Lang}, {Leauthaud}, {Lim}, {Lust}, {MacArthur}, {Mandelbaum}, {Miyatake},
  {Miyazaki}, {Murata}, {More}, {Okura}, {Owen}, {Swinbank}, {Strauss},
  {Yamada}, \& {Yamanoi}}]{Bosch17_hscPipe}
{Bosch}, J., {Armstrong}, R., {Bickerton}, S., {et~al.} 2017, ArXiv e-prints,
  arXiv:1705.06766

\bibitem[{{Bovy}(2015)}]{bovy15}
{Bovy}, J. 2015, \apjs, 216, 29

\bibitem[{{Brandt}(2016)}]{Brandt16}
{Brandt}, T.~D. 2016, \apjl, 824, L31

\bibitem[{{Brown} {et~al.}(2012){Brown}, {Tumlinson}, {Geha}, {Kirby},
  {VandenBerg}, {Mu{\~n}oz}, {Kalirai}, {Simon}, {Avila}, {Guhathakurta},
  {Renzini}, \& {Ferguson}}]{Brown12}
{Brown}, T.~M., {Tumlinson}, J., {Geha}, M., {et~al.} 2012, \apjl, 753, L21

\bibitem[{{Burkert}(1997)}]{Burkert97}
{Burkert}, A. 1997, \apjl, 474, L99

\bibitem[{{Carlin} {et~al.}(2016){Carlin}, {Sand}, {Price}, {Willman},
  {Karunakaran}, {Spekkens}, {Bell}, {Brodie}, {Crnojevi{\'c}}, {Forbes},
  {Hargis}, {Kirby}, {Lupton}, {Peter}, {Romanowsky}, \& {Strader}}]{Carlin16}
{Carlin}, J.~L., {Sand}, D.~J., {Price}, P., {et~al.} 2016, \apjl, 828, L5

\bibitem[{{Carretta} {et~al.}(2009){Carretta}, {Bragaglia}, {Gratton},
  {Lucatello}, {Catanzaro}, {Leone}, {Bellazzini}, {Claudi}, {D'Orazi},
  {Momany}, {Ortolani}, {Pancino}, {Piotto}, {Recio-Blanco}, \&
  {Sabbi}}]{Carretta09}
{Carretta}, E., {Bragaglia}, A., {Gratton}, R.~G., {et~al.} 2009, \aap, 505,
  117

\bibitem[{{Chambers} {et~al.}(2016){Chambers}, {Magnier}, {Metcalfe},
  {Flewelling}, {Huber}, {Waters}, {Denneau}, {Draper}, {Farrow}, {Finkbeiner},
  {Holmberg}, {Koppenhoefer}, {Price}, {Saglia}, {Schlafly}, {Smartt},
  {Sweeney}, {Wainscoat}, {Burgett}, {Grav}, {Heasley}, {Hodapp}, {Jedicke},
  {Kaiser}, {Kudritzki}, {Luppino}, {Lupton}, {Monet}, {Morgan}, {Onaka},
  {Stubbs}, {Tonry}, {Banados}, {Bell}, {Bender}, {Bernard}, {Botticella},
  {Casertano}, {Chastel}, {Chen}, {Chen}, {Cole}, {Deacon}, {Frenk},
  {Fitzsimmons}, {Gezari}, {Goessl}, {Goggia}, {Goldman}, {Grebel}, {Hambly},
  {Hasinger}, {Heavens}, {Heckman}, {Henderson}, {Henning}, {Holman}, {Hopp},
  {Ip}, {Isani}, {Keyes}, {Koekemoer}, {Kotak}, {Long}, {Lucey}, {Liu},
  {Martin}, {McLean}, {Morganson}, {Murphy}, {Nieto-Santisteban}, {Norberg},
  {Peacock}, {Pier}, {Postman}, {Primak}, {Rae}, {Rest}, {Riess}, {Riffeser},
  {Rix}, {Roser}, {Schilbach}, {Schultz}, {Scolnic}, {Szalay}, {Seitz},
  {Shiao}, {Small}, {Smith}, {Soderblom}, {Taylor}, {Thakar}, {Thiel},
  {Thilker}, {Urata}, {Valenti}, {Walter}, {Watters}, {Werner}, {White},
  {Wood-Vasey}, \& {Wyse}}]{ChambersPS1}
{Chambers}, K.~C., {Magnier}, E.~A., {Metcalfe}, N., {et~al.} 2016, ArXiv
  e-prints, arXiv:1612.05560

\bibitem[{{Contenta} {et~al.}(2017{\natexlab{a}}){Contenta}, {Gieles},
  {Balbinot}, \& {Collins}}]{Contenta17}
{Contenta}, F., {Gieles}, M., {Balbinot}, E., \& {Collins}, M.~L.~M.
  2017{\natexlab{a}}, \mnras, 466, 1741

\bibitem[{{Contenta} {et~al.}(2017{\natexlab{b}}){Contenta}, {Balbinot},
  {Petts}, {Read}, {Gieles}, {Collins}, {Pe{\~n}arrubia}, {Delorme}, \&
  {Gualandris}}]{Contenta17_EriII}
{Contenta}, F., {Balbinot}, E., {Petts}, J.~A., {et~al.} 2017{\natexlab{b}},
  ArXiv e-prints, arXiv:1705.01820

\bibitem[{{Crnojevi{\'c}} {et~al.}(2016){Crnojevi{\'c}}, {Sand}, {Zaritsky},
  {Spekkens}, {Willman}, \& {Hargis}}]{Crnojevic16}
{Crnojevi{\'c}}, D., {Sand}, D.~J., {Zaritsky}, D., {et~al.} 2016, \apjl, 824,
  L14

\bibitem[{{Dark Energy Survey Collaboration}(2005)}]{DES05}
{Dark Energy Survey Collaboration}. 2005, ArXiv Astrophysics e-prints,
  astro-ph/0510346

\bibitem[{{Dark Energy Survey Collaboration} {et~al.}(2016){Dark Energy Survey
  Collaboration}, {Abbott}, {Abdalla}, {Aleksi{\'c}}, {Allam}, {Amara},
  {Bacon}, {Balbinot}, {Banerji}, {Bechtol}, {Benoit-L{\'e}vy}, {Bernstein},
  {Bertin}, {Blazek}, {Bonnett}, {Bridle}, {Brooks}, {Brunner}, {Buckley-Geer},
  {Burke}, {Caminha}, {Capozzi}, {Carlsen}, {Carnero-Rosell}, {Carollo},
  {Carrasco-Kind}, {Carretero}, {Castander}, {Clerkin}, {Collett}, {Conselice},
  {Crocce}, {Cunha}, {D'Andrea}, {da Costa}, {Davis}, {Desai}, {Diehl},
  {Dietrich}, {Dodelson}, {Doel}, {Drlica-Wagner}, {Estrada}, {Etherington},
  {Evrard}, {Fabbri}, {Finley}, {Flaugher}, {Foley}, {Fosalba}, {Frieman},
  {Garc{\'{\i}}a-Bellido}, {Gaztanaga}, {Gerdes}, {Giannantonio}, {Goldstein},
  {Gruen}, {Gruendl}, {Guarnieri}, {Gutierrez}, {Hartley}, {Honscheid}, {Jain},
  {James}, {Jeltema}, {Jouvel}, {Kessler}, {King}, {Kirk}, {Kron}, {Kuehn},
  {Kuropatkin}, {Lahav}, {Li}, {Lima}, {Lin}, {Maia}, {Makler}, {Manera},
  {Maraston}, {Marshall}, {Martini}, {McMahon}, {Melchior}, {Merson}, {Miller},
  {Miquel}, {Mohr}, {Morice-Atkinson}, {Naidoo}, {Neilsen}, {Nichol}, {Nord},
  {Ogando}, {Ostrovski}, {Palmese}, {Papadopoulos}, {Peiris}, {Peoples},
  {Percival}, {Plazas}, {Reed}, {Refregier}, {Romer}, {Roodman}, {Ross},
  {Rozo}, {Rykoff}, {Sadeh}, {Sako}, {S{\'a}nchez}, {Sanchez}, {Santiago},
  {Scarpine}, {Schubnell}, {Sevilla-Noarbe}, {Sheldon}, {Smith}, {Smith},
  {Soares-Santos}, {Sobreira}, {Soumagnac}, {Suchyta}, {Sullivan}, {Swanson},
  {Tarle}, {Thaler}, {Thomas}, {Thomas}, {Tucker}, {Vieira}, {Vikram},
  {Walker}, {Wechsler}, {Weller}, {Wester}, {Whiteway}, {Wilcox}, {Yanny},
  {Zhang}, \& {Zuntz}}]{DES16}
{Dark Energy Survey Collaboration}, {Abbott}, T., {Abdalla}, F.~B., {et~al.}
  2016, \mnras, 460, 1270

\bibitem[{{Dotter} {et~al.}(2008){Dotter}, {Chaboyer}, {Jevremovi{\'c}},
  {Kostov}, {Baron}, \& {Ferguson}}]{dotter08}
{Dotter}, A., {Chaboyer}, B., {Jevremovi{\'c}}, D., {et~al.} 2008, \apjs, 178,
  89

\bibitem[{{Drlica-Wagner} \& {MagLiteS Team}(2017)}]{Drlica-Wagner17_MagLiteS}
{Drlica-Wagner}, A., \& {MagLiteS Team}. 2017, in APS April Meeting Abstracts

\bibitem[{{Drlica-Wagner} {et~al.}(2015){Drlica-Wagner}, {Bechtol}, {Rykoff},
  {Luque}, {Queiroz}, {Mao}, {Wechsler}, {Simon}, {Santiago}, {Yanny},
  {Balbinot}, {Dodelson}, {Fausti Neto}, {James}, {Li}, {Maia}, {Marshall},
  {Pieres}, {Stringer}, {Walker}, {Abbott}, {Abdalla}, {Allam},
  {Benoit-L{\'e}vy}, {Bernstein}, {Bertin}, {Brooks}, {Buckley-Geer}, {Burke},
  {Carnero Rosell}, {Carrasco Kind}, {Carretero}, {Crocce}, {da Costa},
  {Desai}, {Diehl}, {Dietrich}, {Doel}, {Eifler}, {Evrard}, {Finley},
  {Flaugher}, {Fosalba}, {Frieman}, {Gaztanaga}, {Gerdes}, {Gruen}, {Gruendl},
  {Gutierrez}, {Honscheid}, {Kuehn}, {Kuropatkin}, {Lahav}, {Martini},
  {Miquel}, {Nord}, {Ogando}, {Plazas}, {Reil}, {Roodman}, {Sako}, {Sanchez},
  {Scarpine}, {Schubnell}, {Sevilla-Noarbe}, {Smith}, {Soares-Santos},
  {Sobreira}, {Suchyta}, {Swanson}, {Tarle}, {Tucker}, {Vikram}, {Wester},
  {Zhang}, {Zuntz}, \& {DES Collaboration}}]{dbr+15}
{Drlica-Wagner}, A., {Bechtol}, K., {Rykoff}, E.~S., {et~al.} 2015, \apj, 813,
  109

\bibitem[{{Drlica-Wagner} {et~al.}(2016){Drlica-Wagner}, {Bechtol}, {Allam},
  {Tucker}, {Gruendl}, {Johnson}, {Walker}, {James}, {Nidever}, {Olsen},
  {Wechsler}, {Cioni}, {Conn}, {Kuehn}, {Li}, {Mao}, {Martin}, {Neilsen},
  {Noel}, {Pieres}, {Simon}, {Stringfellow}, {van der Marel}, \&
  {Yanny}}]{Drlica16}
{Drlica-Wagner}, A., {Bechtol}, K., {Allam}, S., {et~al.} 2016, \apjl, 833, L5

\bibitem[{{Flewelling} {et~al.}(2016){Flewelling}, {Magnier}, {Chambers},
  {Heasley}, {Holmberg}, {Huber}, {Sweeney}, {Waters}, {Chen}, {Farrow},
  {Hasinger}, {Henderson}, {Long}, {Metcalfe}, {Nieto-Santisteban}, {Norberg},
  {Saglia}, {Szalay}, {Rest}, {Thakar}, {Tonry}, {Valenti}, {Werner}, {White},
  {Denneau}, {Draper}, {Hodapp}, {Jedicke}, {Kaiser}, {Kudritzki}, {Price},
  {Wainscoat}, {Chastel}, {McClean}, {Postman}, \& {Shiao}}]{Flewelling16}
{Flewelling}, H.~A., {Magnier}, E.~A., {Chambers}, K.~C., {et~al.} 2016, ArXiv
  e-prints, arXiv:1612.05243

\bibitem[{{Forbes} \& {Kroupa}(2011)}]{forbes_kroupa2011}
{Forbes}, D.~A., \& {Kroupa}, P. 2011, \pasa, 28, 77

\bibitem[{{Geha} {et~al.}(2013){Geha}, {Brown}, {Tumlinson}, {Kalirai},
  {Simon}, {Kirby}, {VandenBerg}, {Mu{\~n}oz}, {Avila}, {Guhathakurta}, \&
  {Ferguson}}]{geha13}
{Geha}, M., {Brown}, T.~M., {Tumlinson}, J., {et~al.} 2013, \apj, 771, 29

\bibitem[{{Hernquist}(1990)}]{hernquist90}
{Hernquist}, L. 1990, \apj, 356, 359

\bibitem[{{Huang} {et~al.}(2017){Huang}, {Leauthaud}, {Murata}, {Bosch},
  {Price}, {Lupton}, {Mandelbaum}, {Lackner}, {Bickerton}, {Miyazaki},
  {Coupon}, \& {Tanaka}}]{Huang17}
{Huang}, S., {Leauthaud}, A., {Murata}, R., {et~al.} 2017, ArXiv e-prints,
  arXiv:1705.01599

\bibitem[{Hunter(2007)}]{Matplotlib}
Hunter, J.~D. 2007, Computing In Science \& Engineering, 9, 90

\bibitem[{{Ivezic} {et~al.}(2008){Ivezic}, {Tyson}, {Abel}, {Acosta},
  {Allsman}, {AlSayyad}, {Anderson}, {Andrew}, {Angel}, {Angeli}, {Ansari},
  {Antilogus}, {Arndt}, {Astier}, {Aubourg}, {Axelrod}, {Bard}, {Barr},
  {Barrau}, {Bartlett}, {Bauman}, {Beaumont}, {Becker}, {Becla}, {Beldica},
  {Bellavia}, {Blanc}, {Blandford}, {Bloom}, {Bogart}, {Borne}, {Bosch},
  {Boutigny}, {Brandt}, {Brown}, {Bullock}, {Burchat}, {Burke}, {Cagnoli},
  {Calabrese}, {Chandrasekharan}, {Chesley}, {Cheu}, {Chiang}, {Claver},
  {Connolly}, {Cook}, {Cooray}, {Covey}, {Cribbs}, {Cui}, {Cutri}, {Daubard},
  {Daues}, {Delgado}, {Digel}, {Doherty}, {Dubois}, {Dubois-Felsmann},
  {Durech}, {Eracleous}, {Ferguson}, {Frank}, {Freemon}, {Gangler}, {Gawiser},
  {Geary}, {Gee}, {Geha}, {Gibson}, {Gilmore}, {Glanzman}, {Goodenow},
  {Gressler}, {Gris}, {Guyonnet}, {Hascall}, {Haupt}, {Hernandez}, {Hogan},
  {Huang}, {Huffer}, {Innes}, {Jacoby}, {Jain}, {Jee}, {Jernigan},
  {Jevremovic}, {Johns}, {Jones}, {Juramy-Gilles}, {Juric}, {Kahn}, {Kalirai},
  {Kallivayalil}, {Kalmbach}, {Kantor}, {Kasliwal}, {Kessler}, {Kirkby},
  {Knox}, {Kotov}, {Krabbendam}, {Krughoff}, {Kubanek}, {Kuczewski},
  {Kulkarni}, {Lambert}, {Le Guillou}, {Levine}, {Liang}, {Lim}, {Lintott},
  {Lupton}, {Mahabal}, {Marshall}, {Marshall}, {May}, {McKercher}, {Migliore},
  {Miller}, {Mills}, {Monet}, {Moniez}, {Neill}, {Nief}, {Nomerotski},
  {Nordby}, {O'Connor}, {Oliver}, {Olivier}, {Olsen}, {Ortiz}, {Owen}, {Pain},
  {Peterson}, {Petry}, {Pierfederici}, {Pietrowicz}, {Pike}, {Pinto}, {Plante},
  {Plate}, {Price}, {Prouza}, {Radeka}, {Rajagopal}, {Rasmussen}, {Regnault},
  {Ridgway}, {Ritz}, {Rosing}, {Roucelle}, {Rumore}, {Russo}, {Saha},
  {Sassolas}, {Schalk}, {Schindler}, {Schneider}, {Schumacher}, {Sebag},
  {Sembroski}, {Seppala}, {Shipsey}, {Silvestri}, {Smith}, {Smith}, {Strauss},
  {Stubbs}, {Sweeney}, {Szalay}, {Takacs}, {Thaler}, {Van Berg}, {Vanden Berk},
  {Vetter}, {Virieux}, {Xin}, {Walkowicz}, {Walter}, {Wang}, {Warner},
  {Willman}, {Wittman}, {Wolff}, {Wood-Vasey}, {Yoachim}, {Zhan}, \& {for the
  LSST Collaboration}}]{IvezicLSST08}
{Ivezic}, Z., {Tyson}, J.~A., {Abel}, B., {et~al.} 2008, ArXiv e-prints,
  arXiv:0805.2366

\bibitem[{{Johnston}(1998)}]{Johnston98}
{Johnston}, K.~V. 1998, \apj, 495, 297

\bibitem[{{Juri{\'c}} {et~al.}(2015){Juri{\'c}}, {Kantor}, {Lim}, {Lupton},
  {Dubois-Felsmann}, {Jenness}, {Axelrod}, {Aleksi{\'c}}, {Allsman},
  {AlSayyad}, {Alt}, {Armstrong}, {Basney}, {Becker}, {Becla}, {Bickerton},
  {Biswas}, {Bosch}, {Boutigny}, {Carrasco Kind}, {Ciardi}, {Connolly},
  {Daniel}, {Daues}, {Economou}, {Chiang}, {Fausti}, {Fisher-Levine},
  {Freemon}, {Gee}, {Gris}, {Hernandez}, {Hoblitt}, {Ivezi{\'c}}, {Jammes},
  {Jevremovi{\'c}}, {Jones}, {Bryce Kalmbach}, {Kasliwal}, {Krughoff}, {Lang},
  {Lurie}, {Lust}, {Mullally}, {MacArthur}, {Melchior}, {Moeyens}, {Nidever},
  {Owen}, {Parejko}, {Peterson}, {Petravick}, {Pietrowicz}, {Price}, {Reiss},
  {Shaw}, {Sick}, {Slater}, {Strauss}, {Sullivan}, {Swinbank}, {Van Dyk},
  {Vuj{\v c}i{\'c}}, {Withers}, {Yoachim}, \& {LSST Project}}]{JuricLSSTDM15}
{Juri{\'c}}, M., {Kantor}, J., {Lim}, K., {et~al.} 2015, ArXiv e-prints,
  arXiv:1512.07914

\bibitem[{{Kafle} {et~al.}(2014){Kafle}, {Sharma}, {Lewis}, \&
  {Bland-Hawthorn}}]{Kafle14}
{Kafle}, P.~R., {Sharma}, S., {Lewis}, G.~F., \& {Bland-Hawthorn}, J. 2014,
  \apj, 794, 59

\bibitem[{{Kerp} {et~al.}(2016){Kerp}, {Kalberla}, {Ben Bekhti}, {Fl{\"o}er},
  {Lenz}, \& {Winkel}}]{kerp16}
{Kerp}, J., {Kalberla}, P.~M.~W., {Ben Bekhti}, N., {et~al.} 2016, \aap, 589,
  A120

\bibitem[{{Kim} \& {Jerjen}(2015)}]{Kim15b}
{Kim}, D., \& {Jerjen}, H. 2015, \apjl, 808, L39

\bibitem[{{Kim} {et~al.}(2015){Kim}, {Jerjen}, {Mackey}, {Da Costa}, \&
  {Milone}}]{Kim15a}
{Kim}, D., {Jerjen}, H., {Mackey}, D., {Da Costa}, G.~S., \& {Milone}, A.~P.
  2015, \apjl, 804, L44

\bibitem[{{King}(1962)}]{king62}
{King}, I. 1962, \aj, 67, 471

\bibitem[{{Kirby} {et~al.}(2013){Kirby}, {Cohen}, {Guhathakurta}, {Cheng},
  {Bullock}, \& {Gallazzi}}]{Kirby13}
{Kirby}, E.~N., {Cohen}, J.~G., {Guhathakurta}, P., {et~al.} 2013, \apj, 779,
  102

\bibitem[{{Kirby} {et~al.}(2015){Kirby}, {Cohen}, {Simon}, \&
  {Guhathakurta}}]{kcs+15}
{Kirby}, E.~N., {Cohen}, J.~G., {Simon}, J.~D., \& {Guhathakurta}, P. 2015,
  \apjl, 814, L7

\bibitem[{{Kirby} {et~al.}(2017){Kirby}, {Cohen}, {Simon}, {Guhathakurta},
  {Thygesen}, \& {Duggan}}]{kcs+17}
{Kirby}, E.~N., {Cohen}, J.~G., {Simon}, J.~D., {et~al.} 2017, \apj, 838, 83

\bibitem[{{Koda} {et~al.}(2015){Koda}, {Yagi}, {Yamanoi}, \&
  {Komiyama}}]{koda15}
{Koda}, J., {Yagi}, M., {Yamanoi}, H., \& {Komiyama}, Y. 2015, \apjl, 807, L2

\bibitem[{{Koposov} {et~al.}(2015){Koposov}, {Belokurov}, {Torrealba}, \&
  {Evans}}]{Koposov15}
{Koposov}, S.~E., {Belokurov}, V., {Torrealba}, G., \& {Evans}, N.~W. 2015,
  \apj, 805, 130

\bibitem[{{Laevens} {et~al.}(2015{\natexlab{a}}){Laevens}, {Martin}, {Bernard},
  {Schlafly}, {Sesar}, {Rix}, {Bell}, {Ferguson}, {Slater}, {Sweeney}, {Wyse},
  {Huxor}, {Burgett}, {Chambers}, {Draper}, {Hodapp}, {Kaiser}, {Magnier},
  {Metcalfe}, {Tonry}, {Wainscoat}, \& {Waters}}]{Laevens15}
{Laevens}, B.~P.~M., {Martin}, N.~F., {Bernard}, E.~J., {et~al.}
  2015{\natexlab{a}}, \apj, 813, 44

\bibitem[{{Laevens} {et~al.}(2015{\natexlab{b}}){Laevens}, {Martin}, {Ibata},
  {Rix}, {Bernard}, {Bell}, {Sesar}, {Ferguson}, {Schlafly}, {Slater},
  {Burgett}, {Chambers}, {Flewelling}, {Hodapp}, {Kaiser}, {Kudritzki},
  {Lupton}, {Magnier}, {Metcalfe}, {Morgan}, {Price}, {Tonry}, {Wainscoat}, \&
  {Waters}}]{lmi+15}
{Laevens}, B.~P.~M., {Martin}, N.~F., {Ibata}, R.~A., {et~al.}
  2015{\natexlab{b}}, \apjl, 802, L18

\bibitem[{{Landsman}(1993)}]{IDLastro}
{Landsman}, W.~B. 1993, in Astronomical Society of the Pacific Conference
  Series, Vol.~52, Astronomical Data Analysis Software and Systems II, ed.
  R.~J. {Hanisch}, R.~J.~V. {Brissenden}, \& J.~{Barnes}, 246

\bibitem[{{{\L}okas} {et~al.}(2013){{\L}okas}, {Gajda}, \&
  {Kazantzidis}}]{Lokas13}
{{\L}okas}, E.~L., {Gajda}, G., \& {Kazantzidis}, S. 2013, \mnras, 433, 878

\bibitem[{{LSST Science Collaboration} {et~al.}(2009){LSST Science
  Collaboration}, {Abell}, {Allison}, {Anderson}, {Andrew}, {Angel}, {Armus},
  {Arnett}, {Asztalos}, {Axelrod}, \& et~al.}]{LSSTScienceBook}
{LSST Science Collaboration}, {Abell}, P.~A., {Allison}, J., {et~al.} 2009,
  ArXiv e-prints, arXiv:0912.0201

\bibitem[{{Magnier} {et~al.}(2016){Magnier}, {Schlafly}, {Finkbeiner}, {Tonry},
  {Goldman}, {R{\"o}ser}, {Schilbach}, {Chambers}, {Flewelling}, {Huber},
  {Price}, {Sweeney}, {Waters}, {Denneau}, {Draper}, {Hodapp}, {Jedicke},
  {Kudritzki}, {Metcalfe}, {Stubbs}, \& {Wainscoast}}]{Magnier16}
{Magnier}, E.~A., {Schlafly}, E.~F., {Finkbeiner}, D.~P., {et~al.} 2016, ArXiv
  e-prints, arXiv:1612.05242

\bibitem[{{Martin} {et~al.}(2008){Martin}, {de Jong}, \& {Rix}}]{mdr08}
{Martin}, N.~F., {de Jong}, J.~T.~A., \& {Rix}, H.-W. 2008, \apj, 684, 1075

\bibitem[{{Martin} {et~al.}(2015){Martin}, {Nidever}, {Besla}, {Olsen},
  {Walker}, {Vivas}, {Gruendl}, {Kaleida}, {Mu{\~n}oz}, {Blum}, {Saha}, {Conn},
  {Bell}, {Chu}, {Cioni}, {de Boer}, {Gallart}, {Jin}, {Kunder}, {Majewski},
  {Martinez-Delgado}, {Monachesi}, {Monelli}, {Monteagudo}, {No{\"e}l},
  {Olszewski}, {Stringfellow}, {van der Marel}, \& {Zaritsky}}]{Martin15}
{Martin}, N.~F., {Nidever}, D.~L., {Besla}, G., {et~al.} 2015, \apjl, 804, L5

\bibitem[{{Martin} {et~al.}(2016){Martin}, {Ibata}, {Collins}, {Rich}, {Bell},
  {Ferguson}, {Laevens}, {Rix}, {Chapman}, \& {Koch}}]{mic+16}
{Martin}, N.~F., {Ibata}, R.~A., {Collins}, M.~L.~M., {et~al.} 2016, \apj, 818,
  40

\bibitem[{{McClure-Griffiths} {et~al.}(2009){McClure-Griffiths}, {Pisano},
  {Calabretta}, {Ford}, {Lockman}, {Staveley-Smith}, {Kalberla}, {Bailin},
  {Dedes}, {Janowiecki}, {Gibson}, {Murphy}, {Nakanishi}, \&
  {Newton-McGee}}]{mcclure09}
{McClure-Griffiths}, N.~M., {Pisano}, D.~J., {Calabretta}, M.~R., {et~al.}
  2009, \apjs, 181, 398

\bibitem[{{McConnachie}(2012)}]{McConnachie12}
{McConnachie}, A.~W. 2012, \aj, 144, 4

\bibitem[{{McMillan}(2017)}]{mcmillan17}
{McMillan}, P.~J. 2017, \mnras, 465, 76

\bibitem[{{Miyamoto} \& {Nagai}(1975)}]{miyamoto-nagai75}
{Miyamoto}, M., \& {Nagai}, R. 1975, \pasj, 27, 533

\bibitem[{{Miyazaki} {et~al.}(2012){Miyazaki}, {Komiyama}, {Nakaya}, {Kamata},
  {Doi}, {Hamana}, {Karoji}, {Furusawa}, {Kawanomoto}, {Morokuma}, {Ishizuka},
  {Nariai}, {Tanaka}, {Uraguchi}, {Utsumi}, {Obuchi}, {Okura}, {Oguri},
  {Takata}, {Tomono}, {Kurakami}, {Namikawa}, {Usuda}, {Yamanoi}, {Terai},
  {Uekiyo}, {Yamada}, {Koike}, {Aihara}, {Fujimori}, {Mineo}, {Miyatake},
  {Yasuda}, {Nishizawa}, {Saito}, {Tanaka}, {Uchida}, {Katayama}, {Wang},
  {Chen}, {Lupton}, {Loomis}, {Bickerton}, {Price}, {Gunn}, {Suzuki},
  {Miyazaki}, {Muramatsu}, {Yamamoto}, {Endo}, {Ezaki}, {Itoh}, {Miwa},
  {Yokota}, {Matsuda}, {Ebinuma}, \& {Takeshi}}]{HSC}
{Miyazaki}, S., {Komiyama}, Y., {Nakaya}, H., {et~al.} 2012, in \procspie, Vol.
  8446, Ground-based and Airborne Instrumentation for Astronomy IV, 84460Z

\bibitem[{{Mu{\~n}oz} {et~al.}(2006){Mu{\~n}oz}, {Carlin}, {Frinchaboy},
  {Nidever}, {Majewski}, \& {Patterson}}]{Munoz06_Bootes}
{Mu{\~n}oz}, R.~R., {Carlin}, J.~L., {Frinchaboy}, P.~M., {et~al.} 2006, \apjl,
  650, L51

\bibitem[{{Mu{\~n}oz} {et~al.}(2010){Mu{\~n}oz}, {Geha}, \&
  {Willman}}]{Munoz10}
{Mu{\~n}oz}, R.~R., {Geha}, M., \& {Willman}, B. 2010, \aj, 140, 138

\bibitem[{{Mu{\~n}oz} {et~al.}(2012){Mu{\~n}oz}, {Padmanabhan}, \&
  {Geha}}]{Munoz12}
{Mu{\~n}oz}, R.~R., {Padmanabhan}, N., \& {Geha}, M. 2012, \apj, 745, 127

\bibitem[{{Nidever} {et~al.}(2017){Nidever}, {Olsen}, {Walker}, {Vivas},
  {Blum}, {Kaleida}, {Choi}, {Conn}, {Gruendl}, {Bell}, {Besla}, {Munoz},
  {Gallart}, {Martin}, {Olszewski}, {Saha}, {Monachesi}, {Monelli}, {de Boer},
  {Johnson}, {Zaritsky}, {Stringfellow}, {van der Marel}, {Cioni}, {Jin},
  {Majewski}, {Martinez-Delgado}, {Monteagudo}, {Noel}, {Bernard}, {Kunder},
  {Chu}, {Bell}, {Santana}, {Frechem}, {Medina}, {Parkash}, {Seron}, \&
  {Hayes}}]{Nidever17_SMASH}
{Nidever}, D.~L., {Olsen}, K., {Walker}, A.~R., {et~al.} 2017, ArXiv e-prints,
  arXiv:1701.00502

\bibitem[{{Oh} {et~al.}(1995){Oh}, {Lin}, \& {Aarseth}}]{OhLinAarseth95}
{Oh}, K.~S., {Lin}, D.~N.~C., \& {Aarseth}, S.~J. 1995, \apj, 442, 142

\bibitem[{{Pe{\~n}arrubia} {et~al.}(2008){Pe{\~n}arrubia}, {Navarro}, \&
  {McConnachie}}]{penarrubia08}
{Pe{\~n}arrubia}, J., {Navarro}, J.~F., \& {McConnachie}, A.~W. 2008, \apj,
  673, 226

\bibitem[{Perez \& Granger(2007)}]{Perez07_iPython}
Perez, F., \& Granger, B.~E. 2007, Computing in Science Engineering, 9, 21

\bibitem[{{Perryman} {et~al.}(2001){Perryman}, {de Boer}, {Gilmore}, {H{\o}g},
  {Lattanzi}, {Lindegren}, {Luri}, {Mignard}, {Pace}, \& {de
  Zeeuw}}]{Perryman01}
{Perryman}, M.~A.~C., {de Boer}, K.~S., {Gilmore}, G., {et~al.} 2001, \aap,
  369, 339

\bibitem[{{Rockosi} {et~al.}(2002){Rockosi}, {Odenkirchen}, {Grebel}, {Dehnen},
  {Cudworth}, {Gunn}, {York}, {Brinkmann}, {Hennessy}, \&
  {Ivezi{\'c}}}]{rog+02}
{Rockosi}, C.~M., {Odenkirchen}, M., {Grebel}, E.~K., {et~al.} 2002, \aj, 124,
  349

\bibitem[{{Sand} {et~al.}(2009){Sand}, {Olszewski}, {Willman}, {Zaritsky},
  {Seth}, {Harris}, {Piatek}, \& {Saha}}]{Sand09}
{Sand}, D.~J., {Olszewski}, E.~W., {Willman}, B., {et~al.} 2009, \apj, 704, 898

\bibitem[{{Sand} {et~al.}(2012){Sand}, {Strader}, {Willman}, {Zaritsky},
  {McLeod}, {Caldwell}, {Seth}, \& {Olszewski}}]{Sand12}
{Sand}, D.~J., {Strader}, J., {Willman}, B., {et~al.} 2012, \apj, 756, 79

\bibitem[{{Schlafly} \& {Finkbeiner}(2011)}]{schlafly11}
{Schlafly}, E.~F., \& {Finkbeiner}, D.~P. 2011, \apj, 737, 103

\bibitem[{{Schlegel} {et~al.}(1998){Schlegel}, {Finkbeiner}, \&
  {Davis}}]{SFD98}
{Schlegel}, D.~J., {Finkbeiner}, D.~P., \& {Davis}, M. 1998, \apj, 500, 525

\bibitem[{{Shanks} {et~al.}(2015){Shanks}, {Metcalfe}, {Chehade}, {Findlay},
  {Irwin}, {Gonzalez-Solares}, {Lewis}, {Yoldas}, {Mann}, {Read}, {Sutorius},
  \& {Voutsinas}}]{Shanks15_ATLAS}
{Shanks}, T., {Metcalfe}, N., {Chehade}, B., {et~al.} 2015, \mnras, 451, 4238

\bibitem[{{Simon} \& {Geha}(2007)}]{Simon07}
{Simon}, J.~D., \& {Geha}, M. 2007, \apj, 670, 313

\bibitem[{{Spekkens} {et~al.}(2014){Spekkens}, {Urbancic}, {Mason}, {Willman},
  \& {Aguirre}}]{Spekkens14}
{Spekkens}, K., {Urbancic}, N., {Mason}, B.~S., {Willman}, B., \& {Aguirre},
  J.~E. 2014, \apjl, 795, L5

\bibitem[{{Taylor}(2005)}]{Topcat}
{Taylor}, M.~B. 2005, in Astronomical Society of the Pacific Conference Series,
  Vol. 347, Astronomical Data Analysis Software and Systems XIV, ed.
  P.~{Shopbell}, M.~{Britton}, \& R.~{Ebert}, 29

\bibitem[{{Torrealba} {et~al.}(2016){Torrealba}, {Koposov}, {Belokurov}, \&
  {Irwin}}]{Torrealba16}
{Torrealba}, G., {Koposov}, S.~E., {Belokurov}, V., \& {Irwin}, M. 2016,
  \mnras, 459, 2370

\bibitem[{van~der Walt {et~al.}(2011)van~der Walt, Colbert, \&
  Varoquaux}]{vanderWalt11_numpy}
van~der Walt, S., Colbert, S.~C., \& Varoquaux, G. 2011, Computing in Science
  Engineering, 13, 22

\bibitem[{{van Dokkum} {et~al.}(2016){van Dokkum}, {Abraham}, {Brodie},
  {Conroy}, {Danieli}, {Merritt}, {Mowla}, {Romanowsky}, \&
  {Zhang}}]{vanDokkum16}
{van Dokkum}, P., {Abraham}, R., {Brodie}, J., {et~al.} 2016, \apjl, 828, L6

\bibitem[{{Venn} {et~al.}(2017){Venn}, {Starkenburg}, {Malo}, {Martin}, \&
  {Laevens}}]{vsm+17}
{Venn}, K.~A., {Starkenburg}, E., {Malo}, L., {Martin}, N., \& {Laevens},
  B.~P.~M. 2017, \mnras, 466, 3741

\bibitem[{{Wakker} \& {van Woerden}(1997)}]{WvW97}
{Wakker}, B.~P., \& {van Woerden}, H. 1997, \araa, 35, 217

\bibitem[{{Walsh} {et~al.}(2007){Walsh}, {Jerjen}, \&
  {Willman}}]{Walsh07_BooII}
{Walsh}, S.~M., {Jerjen}, H., \& {Willman}, B. 2007, \apjl, 662, L83

\bibitem[{{Willman} \& {Strader}(2012)}]{WillmanStrader12}
{Willman}, B., \& {Strader}, J. 2012, \aj, 144, 76

\bibitem[{{Willman} {et~al.}(2005){Willman}, {Dalcanton}, {Martinez-Delgado},
  {West}, {Blanton}, {Hogg}, {Barentine}, {Brewington}, {Harvanek}, {Kleinman},
  {Krzesinski}, {Long}, {Neilsen}, {Nitta}, \& {Snedden}}]{Willman05_UMa}
{Willman}, B., {Dalcanton}, J.~J., {Martinez-Delgado}, D., {et~al.} 2005,
  \apjl, 626, L85

\bibitem[{{Winkel} {et~al.}(2016){Winkel}, {Kerp}, {Fl{\"o}er}, {Kalberla},
  {Ben Bekhti}, {Keller}, \& {Lenz}}]{winkel16}
{Winkel}, B., {Kerp}, J., {Fl{\"o}er}, L., {et~al.} 2016, \aap, 585, A41

\bibitem[{{Wolf} {et~al.}(2010){Wolf}, {Martinez}, {Bullock}, {Kaplinghat},
  {Geha}, {Mu{\~n}oz}, {Simon}, \& {Avedo}}]{wolf10}
{Wolf}, J., {Martinez}, G.~D., {Bullock}, J.~S., {et~al.} 2010, \mnras, 406,
  1220

\bibitem[{{York} {et~al.}(2000){York}, {Adelman}, {Anderson}, {Anderson},
  {Annis}, {Bahcall}, {Bakken}, {Barkhouser}, {Bastian}, {Berman}, {Boroski},
  {Bracker}, {Briegel}, {Briggs}, {Brinkmann}, {Brunner}, {Burles}, {Carey},
  {Carr}, {Castander}, {Chen}, {Colestock}, {Connolly}, {Crocker}, {Csabai},
  {Czarapata}, {Davis}, {Doi}, {Dombeck}, {Eisenstein}, {Ellman}, {Elms},
  {Evans}, {Fan}, {Federwitz}, {Fiscelli}, {Friedman}, {Frieman}, {Fukugita},
  {Gillespie}, {Gunn}, {Gurbani}, {de Haas}, {Haldeman}, {Harris}, {Hayes},
  {Heckman}, {Hennessy}, {Hindsley}, {Holm}, {Holmgren}, {Huang}, {Hull},
  {Husby}, {Ichikawa}, {Ichikawa}, {Ivezi{\'c}}, {Kent}, {Kim}, {Kinney},
  {Klaene}, {Kleinman}, {Kleinman}, {Knapp}, {Korienek}, {Kron}, {Kunszt},
  {Lamb}, {Lee}, {Leger}, {Limmongkol}, {Lindenmeyer}, {Long}, {Loomis},
  {Loveday}, {Lucinio}, {Lupton}, {MacKinnon}, {Mannery}, {Mantsch}, {Margon},
  {McGehee}, {McKay}, {Meiksin}, {Merelli}, {Monet}, {Munn}, {Narayanan},
  {Nash}, {Neilsen}, {Neswold}, {Newberg}, {Nichol}, {Nicinski}, {Nonino},
  {Okada}, {Okamura}, {Ostriker}, {Owen}, {Pauls}, {Peoples}, {Peterson},
  {Petravick}, {Pier}, {Pope}, {Pordes}, {Prosapio}, {Rechenmacher}, {Quinn},
  {Richards}, {Richmond}, {Rivetta}, {Rockosi}, {Ruthmansdorfer}, {Sandford},
  {Schlegel}, {Schneider}, {Sekiguchi}, {Sergey}, {Shimasaku}, {Siegmund},
  {Smee}, {Smith}, {Snedden}, {Stone}, {Stoughton}, {Strauss}, {Stubbs},
  {SubbaRao}, {Szalay}, {Szapudi}, {Szokoly}, {Thakar}, {Tremonti}, {Tucker},
  {Uomoto}, {Vanden Berk}, {Vogeley}, {Waddell}, {Wang}, {Watanabe},
  {Weinberg}, {Yanny}, {Yasuda}, \& {SDSS Collaboration}}]{York2000_SDSS}
{York}, D.~G., {Adelman}, J., {Anderson}, Jr., J.~E., {et~al.} 2000, \aj, 120,
  1579

\bibitem[{{Zaritsky}(2017)}]{zaritsky17}
{Zaritsky}, D. 2017, \mnras, 464, L110

\bibitem[{{Zucker} {et~al.}(2006{\natexlab{a}}){Zucker}, {Belokurov}, {Evans},
  {Kleyna}, {Irwin}, {Wilkinson}, {Fellhauer}, {Bramich}, {Gilmore}, {Newberg},
  {Yanny}, {Smith}, {Hewett}, {Bell}, {Rix}, {Gnedin}, {Vidrih}, {Wyse},
  {Willman}, {Grebel}, {Schneider}, {Beers}, {Kniazev}, {Barentine},
  {Brewington}, {Brinkmann}, {Harvanek}, {Kleinman}, {Krzesinski}, {Long},
  {Nitta}, \& {Snedden}}]{Zucker06_Uma}
{Zucker}, D.~B., {Belokurov}, V., {Evans}, N.~W., {et~al.} 2006{\natexlab{a}},
  \apjl, 650, L41

\bibitem[{{Zucker} {et~al.}(2006{\natexlab{b}}){Zucker}, {Belokurov}, {Evans},
  {Wilkinson}, {Irwin}, {Sivarani}, {Hodgkin}, {Bramich}, {Irwin}, {Gilmore},
  {Willman}, {Vidrih}, {Fellhauer}, {Hewett}, {Beers}, {Bell}, {Grebel},
  {Schneider}, {Newberg}, {Wyse}, {Rockosi}, {Yanny}, {Lupton}, {Smith},
  {Barentine}, {Brewington}, {Brinkmann}, {Harvanek}, {Kleinman}, {Krzesinski},
  {Long}, {Nitta}, \& {Snedden}}]{Zucker06_CVn}
---. 2006{\natexlab{b}}, \apjl, 643, L103

\end{thebibliography}

\end{document}